\makeatletter
\def\@sect#1#2#3#4#5#6[#7]#8{\ifnum #2>\c@secnumdepth
     \def\@svsec{}\else
     \refstepcounter{#1}\edef\@svsec{\csname the#1\endcsname.\hskip 1em }\fi
     \@tempskipa #5\relax
      \ifdim \@tempskipa>\z@
        \begingroup #6\relax
          \@hangfrom{\hskip #3\relax\@svsec}{\interlinepenalty \@M #8\par}
        \endgroup
       \csname #1mark\endcsname{#7}\addcontentsline
         {toc}{#1}{\ifnum #2>\c@secnumdepth \else
                      \protect\numberline{\csname the#1\endcsname}\fi
                    #7}\else
        \def\@svsechd{#6\hskip #3\@svsec #8\csname #1mark\endcsname
                      {#7}\addcontentsline
                           {toc}{#1}{\ifnum #2>\c@secnumdepth \else
                             \protect\numberline{\csname the#1\endcsname}\fi
                       #7}}\fi
     \@xsect{#5}}
\def\label#1{\@bsphack\if@filesw {\let\thepage\relax
   \xdef\@gtempa{\write\@auxout{\string
      \newlabel{#1}{{\thesection.\@currentlabel}{\thepage}}}}}\@gtempa
   \if@nobreak \ifvmode\nobreak\fi\fi\fi\@esphack}
\def\@eqnnum{(\thesection.\theequation)}
\documentstyle[12pt,citesort]{article}
\makeatletter
\def\section{\setcounter{equation}{0} \@startsection {section}{1}{\z@}{-3.5ex
   plus -1ex minus -.2ex}{2.3ex plus .2ex}{\Large\bf}}
\def\IJMP #1 #2 #3 {{\it Int.\ J.\ Mod.\ Phys.}\ {\bf #1}\ (#2) #3}
\def\MPL #1 #2 #3 {{\it Mod.\ Phys.\ Lett.}\ {\bf #1}\ (#2) #3}
\def\NC #1 #2 #3 {{\it Nuovo Cim.}\ {\bf #1} (#2) #3}
\def\NP #1 #2 #3 {{\it Nucl.\ Phys.}\ {\bf #1}\ (#2) #3}
\def\PL #1 #2 #3 {{\it Phys.\ Lett.}\ {\bf #1}\ (#2) #3}
\def\PR #1 #2 #3 {{\it Phys.\ Rev.}\ {\bf #1}\ (#2) #3}
\def\PP #1 #2 #3 {{\it Phys.\ Rep.}\ {\bf #1}\ (#2) #3}
\def\PRL #1 #2 #3 {{\it Phys.\ Rev.\ Lett.}\ {\bf #1}\ (#2) #3}
\def\RMP #1 #2 #3 {{\it Rev.\ Mod.\ Phys.}\ {\bf #1}\ (#2) #3}
\def\ZP #1 #2 #3 {{\it Z.\ Phys.}\ {\bf #1}\ (#2) #3}
\def\g{\gamma}
\date{}
\begin{document}
\begin{flushright}
IHEP 94--77\\
hep-ph/9407393\\
July 1994
\end{flushright}

\vspace*{0.5in}
\begin{center}
{\bf 	FOUR WEAK GAUGE BOSON PRODUCTION\\
	AT PHOTON LINEAR COLLIDER\\
	AND HEAVY HIGGS SIGNAL}\\
[1ex]	G. Jikia\\
[1ex]	{\it Institute for High Energy Physics} \\
	{\it 142284, Protvino, Moscow region, Russia}
\end{center}
\begin{abstract}
We study the signals and backgrounds for a heavy Higgs boson in the
processes $\gamma\gamma\to WWWW$, $\gamma\gamma\to WWZZ$ at the photon
linear collider. The results are based on the complete tree level SM
calculation for these reactions. We show that the invariant mass spectrum of
central $WW$, $ZZ$ pairs is sensitive to the signal from Higgs boson with a
mass up to 1~TeV at a 2~TeV linear collider for integrated luminosity of
300~fb$^{-1}$. At 1.5~TeV PLC Higgs boson with a mass up to 700~GeV can be
studied. The nonresonant longitudinal gauge boson scattering ($m_H=\infty$)
can be detected in photon-photon collisions at $e^+e^-$ center-of-mass
energy of 3~TeV.
\end{abstract}

\section{Introduction}

One of the most challenging puzzles of contemporary particle physics is
whether Nature indeed makes use of the Higgs mechanism of spontaneous
electroweak symmetry breaking. If Higgs boson will be found below 800~GeV or
so, this will be a proof of the so called weak scenario of the symmetry
breaking (SB). Otherwise the scenario of the strongly interacting
electroweak SB (EWSB) will take place (for recent reviews see, {\it
e.g.}, \cite{SB}). The study of strong EWSB is one of the major motivations
to built the next generation of colliders.  While the potential of hadronic
colliders (see, {\it e.g.}, \cite{Gold} and references therein) as well as
linear $e^+e^-$ colliders \cite{SB,hagiwara,kurihara} to explore EWSB was
extensively studied, much less was done for $\gamma\gamma$ colliders
\cite{gg,gg1,gg2,gg3}, which would provide additional unique capabilities
\cite{brodsky-talk,chanowitz}.

The would be ``gold-plated" channel for Higgs boson production at
a Photon Linear Collider (PLC)
\begin{equation}
\gamma\gamma\to H\to ZZ\to (q\bar q) (l^+l^-)
\end{equation}
was shown recently to be suffered from very large background from continuum
$ZZ$ pair production through $W$ boson loop for the Higgs mass above 350~GeV
\cite{aazz}. So this reaction, although very promising for the measurement
of the two-photon Higgs width for $M_H\leq 300-400$~GeV, provides very poor
possibilities for studying of a heavy Higgs and EWSB
\cite{rosenfeld,berger-chanowitz}, unless there are
strong tensor resonances within the energy reach of PLC
\cite{berger-chanowitz}.

Another very interesting potential application of photon-photon collisions
at a high energy linear collider proposed recently \cite{brodsky,boudjema}
is $WW$ scattering, as illustrated in Fig. 1. In this process each photon is
resolved as a $WW$ pair. The interacting vector bosons can then scatter
pair-wise or annihilate; {\it e.g.}, they can annihilate into a Higgs boson
decaying into a $WW$ or $ZZ$ pair. In principle, one can use these processes
\begin{eqnarray}
\g(1)+ \g(2)&\to& W^+(3)+ W^+(4)+ W^-(5)+ W^-(6), \nonumber\\
\g(1)+ \g(2)&\to& W^+(3)+ W^-(4)+ Z(5)+ Z(6)
\label{WWWW}
\end{eqnarray}
for studying EWSB. In fact, this reaction at photon-photon collider is an
analog of the reaction $e^+e^-\to \nu\bar\nu WW(\nu\bar\nu ZZ)$ at linear
$e^+e^-$ collider.  Event rates for the reaction (\ref{WWWW}) were estimated
using effective $W$ approximation (EWA) \cite{kingman} for several models of
EWSB and quite optimistic conclusions were given. However, EWA  has a
limited accuracy at energies of 1--2~TeV and, moreover, it does not permit
to calculate the effects of the tag and veto cuts used to isolate the Higgs
signal. In fact, much more diagrams (see Fig.~2) contribute to reactions
(\ref{WWWW}) and it is unclear a priori that the background from $WWWW$,
$WWZZ$ final states, all vector bosons being transverse, is manageable.  We
have presented our first results of the exact standard model (SM) tree level
calculation for the reactions (\ref{WWWW}) in \cite{LBL-talk} and
demonstrated that the observation of the heavy Higgs signal above the
background is possible. Subsequently, the background was recalculated and,
in addition, various models of EWSB were considered in \cite{cheung}
confirming calculation \cite{LBL-talk}.

This paper extends our earlier results \cite{LBL-talk} on the heavy Higgs
signal and background based on complete leading order calculation for
reactions (\ref{WWWW}). In Section~2 we compare effective $W_LW_L$
luminosity in photon-photon and $e^+e^-$ collisions. The details of the
calculation are given in Section~3. In Section~4 we present total cross
sections for different polarizations of $WWWW$, $WWZZ$ final states. In
Section~5 we will concentrate on the heavy ($m_H=1$~TeV) SM Higgs boson case
as a prototype for models of strong EWSB and will show that its signal can
be observed at a 2~TeV linear collider. We also show that Higgs boson with a
mass up to 700~GeV should be relatively easily observed in photon-photon
collisions at a 1.5~TeV linear collider. We conclude with some brief remarks
in Section~6.

\section{Effective $W_LW_L$ luminosity in photon-photon collisions}

In this section we try to estimate the $W_L$ flux in photon-photon
collisions in comparison to that in $e^+e^-$ collisions before calculating
$\g\g\to WWZZ,\,WWWW$ cross sections. An expression for the longitudinal $W$
content inside the photon has been derived \cite{W/gamma}
\begin{equation}
f_{W_L/\g}(z) = \frac{\alpha}{\pi} \left[\frac{1-z}{z}+\frac{z(1-z)}{2}
\left(log\frac{s(1-z)^2}{M_W^2}-2\right)\right].
\label{W/gamma}
\end{equation}
The conclusion was made that the photon had a more abundant distribution
function at large $z$ than electron \cite{W/e}
\begin{equation}
f_{W_L/e}(z) = \frac{\alpha}{4\pi\sin^2\theta_W} \frac{1-z}{z}
\label{W/e}
\end{equation}
due to the hard component with the logarithmic enhancement factor. The
derivation of polarized distribution functions of quasi-real equivalent
$W_L$ inside the photon is quite a subtle problem (see, {\it e.g.},
discussion in \cite{boudjema}), so we will not rely on expressions like
(\ref{W/gamma}) and (\ref{W/e}) here. We just define the effective $W_LW_L$
luminosity in photon-photon collisions as
\begin{equation}
\sigma(\g\g\to W^+W^-H)\,\equiv\,16\pi^2\,\frac{\Gamma(H\to W^+W^-)}{m_H^3}
\,\left.\tau \frac{d{\cal L}}{d\tau}\right|_{W_LW_L/\g\g}.
\label{WW/gg}
\end{equation}
Here left hand side is the exact tree-level cross section for $W^+W^-H$
production in photon-photon collisions, $\tau = m_H^2/s = M_{W_LW_L}^2/s$.
It is assumed that Higgs boson mass is large enough, so that transverse
$W_TW_T$ contribution is negligible. The form of the right hand side is
derived from the expression of the cross section of Higgs boson production
in $WW$ fusion in terms of longitudinal $W$ distribution function calculated
relying on equivalent $W_L$ approximation
\begin{equation}
\sigma_{EWA}(H) = 16\pi^2 \frac{\Gamma(H\to W_L^+W_L^-)}{m_H^3}
\tau \int\frac{dz}{z}f_{W_L}(z)f_{W_L}(\frac{\tau}{z}).
\label{EWA}
\end{equation}
We stress again that we will not use the EWA formulas like
(\ref{W/gamma},\ref{W/e}), but instead will use the equation (\ref{WW/gg})
as a definition of ${\cal L}_{W_LW_L/\g\g}$, thereby avoiding inaccuracies
associated with expression (\ref{W/gamma}). We believe this approach extends
the applicability of EWA method to a wide kinematical region and foresees the
use of any next-to-leading improved EWA distribution functions.  Similarly,
we define the effective $W_LW_L$ luminosity in $e^+e^-$ collisions as
\begin{equation}
\sigma(e^+e^-\to \nu_e\bar\nu_e H)\,\equiv\,
16\pi^2\,\frac{\Gamma(H\to W^+W^-)}{m_H^3}\,
\left.\tau \frac{d{\cal L}}{d\tau}\right|_{W_LW_L/e^+e^-}.
\end{equation}

Fig.~3 presents the effective $W_LW_L$ luminosity in $\g\g$ and $e^+e^-$
collisions as a function of $\tau$ for various values of center-of-mass
energies. The $W_LW_L$ luminosity in $e^+e^-$ collisions is almost
independent of $\sqrt{s}$ as a function of $\tau$ at $\tau>0.1$ exhibiting
scaling behaviour, as it is implied by (\ref{W/e}). On the contrary, the
$W_LW_L$ luminosity in $\g\g$ collisions does depend on photon-photon
center-of-mass energy, that again can be expected from EWA distribution
function (\ref{W/gamma}). While at $\sqrt{s}=1$~TeV the $W_LW_L$ luminosity
in photon-photon collisions coincides with the luminosity in $e^+e^-$
collisions at large values of $\tau$ and is slightly smaller at low values
of $\tau$, at higher energies ${\cal L}_{W_LW_L/\g\g}$ is several times
larger than ${\cal L}_{W_LW_L/e^+e^-}$. This confirms previous hopes that at
multi-TeV energies photon beams can be more efficient source of energetic
$W_L$'s than electron beams. In fact, the larger luminosity in photon-photon
collisions results from harder distribution function (\ref{W/gamma}) and an
additional factor of two, because both photons can produce $W_L^+$ as well
as $W_L^-$. However, to compare the potential of photon-photon option of
linear collider with that of original $e^+e^-$ option, one has to take into
account that laser induced photon beams are not monochromatic \cite{gg}:
photon spectrum depends on the product of the helicity of the electron and
laser photon and the highest photon energy is bounded by
\begin{equation}
\omega_{max} = \frac{x_0}{x_0+1}\,E_{beam},
\end{equation}
where $x_0$ is the machine parameter related to the electron beam energy and
laser photon energy \cite{gg}
\begin{equation}
x_0 = \frac{4 E_{beam}\omega_0}{m_e^2}.
\end{equation}
The optimal value of $x_0$ is 4.8 \cite{gg}, so that the maximum photon
energy fraction is $z_{max} = 0.8$. A more realistic effective $W_LW_L$
luminosity in photon-photon collisions calculated taking into account the
photon distribution function
\begin{equation}
\tau\,\left.\frac{d{\cal L}^{eff}}{d\tau}(\tau)\right|_{W_LW_L/\g\g}\,=\,
\tau\,\int\frac{dz_1}{z_1}\frac{dz_2}{z_2}f_\g(z_1)f_\g(z_2)
\,\left.\frac{d{\cal L}}{d\tau}(\frac{\tau}{z_1z_2})\right|_{W_LW_L/\g\g}
\end{equation}
is presented in Fig.~4. We assume that the product of electron and laser
helicity is $2\lambda_e\lambda_\g =2\lambda_e'\lambda_\g' = -0.9$. This case
gives the hardest photon spectrum for 100\% polarized laser beams and 90\%
polarized electron beams. For initial laser beam helicities of $++$ and
$+-$, final high energy photons are produced predominantly in $J_Z=0$ and
$J_Z=2$ states, respectively. The effective luminosity is about two times
larger for $++$ initial laser helicities, but it is more than an order of
magnitude smaller than for monochromatic photons. And for all energies the
$W_LW_L$ luminosity in photon-photon collisions is now almost an order of
magnitude smaller than the $W_LW_L$ luminosity in $e^+e^-$ collisions at the
same center-of-mass energy. This fact can be easily understood. Consider,
for example, the $W_LW_L$ luminosity for 2~TeV linear collider at
$M_{W_LW_L}=1$~TeV. The ${\cal L}_{W_LW_L/\g\g}$ is two times larger than
${\cal L}_{W_LW_L/e^+e^-}$ at $\sqrt{s_{\g\g}}=\sqrt{s_{e^+e^-}}=2$~TeV.
However, the maximum photon-photon center-of-mass energy is $z_{max}$ times
smaller that $e^+e^-$ energy. The $W_LW_L$ luminosity in photon-photon
collisions steeply rises as a function of $\sqrt{s_{\g\g}}$ (see Fig.~5). At
1.6~TeV ${\cal L}_{W_LW_L/\g\g}$ is 3 times smaller than at
$\sqrt{s_{\g\g}}=2$~TeV. Moreover, the integral over photon-photon
luminosity spectrum in the interval
$(M_{W_LW_L}+2M_W)/\sqrt{s}=0.6<\sqrt{\tau}<0.8=z_{max}$ is 0.35. And,
finally, a factor of 2 is lost because within this interval $W_LW_L$
luminosity decreases when $\tau$ decreases. Thus, at $\sqrt{s}=2$~TeV and
$M_{W_LW_L}=1$~TeV effective $W_LW_L$ luminosity in photon-photon collisions
for realistic photon spectrum is 17 times smaller than that for
monochromatic photon spectrum and 8 times smaller than $W_LW_L$ luminosity
in $e^+e^-$ collisions.

Nevertheless, what should be emphasized, is that the luminosity of high
energy photon-photon collider has a much less restrictive upper bound than
that for $e^+e^-$ collider because of different conditions at the
interaction point  \cite{gg1,gg2,gg3}. And it is even stated that such a
huge luminosity as $10^{35-36}$~cm$^{-2}$~s$^{-1}$ might be achievable in
photon-photon collisions \cite{gg1,gg2,gg3}. So, it is quite possible that
the lack of $W_LW_L$ luminosity due to non-monochromaticity of the laser
induced photon spectrum can be compensated by technical advantages of the
photon-photon option of the linear collider.

\section{Matrix element}

The calculation of four weak gauge boson production in photon-photon
collisions involves a large number of Feynman diagrams and, so, the
efficiency and numerical stability are key issues. {\it E.g.}, if one would
calculate the square of the matrix element in unitary gauge, severe
numerical cancellations would occur among the longitudinal $p_\mu
p_\nu/M_W^2$ terms, as the cross section should be non-singular in the
limit $M_W\to 0$. At $\sqrt{s}=2$~TeV the matrix element squared of $WWWW$,
$WWZZ$ production will contain the most singular terms of the order of
$(E_W/M_W)^{20}\sim 10^{16}$, {\it i.e.} sixteen decimal digits will be lost!
Hence, it is advantageous to use renormalizable gauge, {\it
e.g.}, 't~Hooft-Feynman gauge, and work with amplitudes directly and square
them numerically. Then the most singular amplitude of four longitudinal gauge
boson production will involve cancellation of terms of the order of
$(E_W/M_W)^4$, implying a loss of accuracy of three or six digits at
photon-photon center-of-mass energy of 2~TeV or 10~TeV, respectively.
Further, it is possible to reduce the number of diagrams using non-linear
gauges \cite{non-linear} where mixed  photon--$W$--Nambu-Goldstone triple
vertices are absent.

There are five topologically distinctive graphs contributing to tree-level
reaction with six external particles and all the diagrams describing
processes (\ref{WWWW}) can be easily generated. Some Feynman diagrams are
shown in Fig.~2, where also are given multiplicities of diagrams for a given
topology. We used a variant of non-linear gauge which we have exploited
earlier for the calculation of one-loop reaction $\g\g\to ZZ$ \cite{aazz}.
In total, there are 240 diagrams for $WWWW$ production and 104 diagrams for
$WWZZ$ production. In 't~Hooft-Feynman gauge the number of diagrams is
almost three times larger.

We used symbolic manipulation program FORM \cite{FORM} to express the
amplitude in the following form
\begin{eqnarray}
&&{\cal M}(\g\g\to WWWW,\ WWZZ) = \nonumber\\
&&\sum_{i_1\dots i_6,j_1\dots j_4}
A_{i_1\dots i_6,j_1\dots j_4}(e_{i_1}e_{i_2})(e_{i_3}p_{j_1})
(e_{i_4}p_{j_2}) (e_{i_5}p_{j_3}) (e_{i_6}p_{j_4}) \nonumber\\
&&+\sum_{i_1\dots i_6,j_1,j_2}
B_{i_1\dots i_6,j_1,j_2}(e_{i_1}e_{i_2})(e_{i_3}e_{i_4})
(e_{i_5}p_{j_1}) (e_{i_6}p_{j_2}) \nonumber\\
&&+\sum_{i_1\dots i_6}
C_{i_1\dots i_6}(e_{i_1}e_{i_2})(e_{i_3}e_{i_4})
(e_{i_5}e_{i_6}),
\label{ME}
\end{eqnarray}
here $p_{1-6}$ and $e_{1-6}$ are momenta and polarization vectors. This
reduces the amount of numerical work when computing cross section summed
over all helicity states, because coefficients $A$, $B$ and $C$ are
independent of the polarizations and they have to be computed only once. In
fact, all the terms in (\ref{ME}) have been further bracketed in such a way,
that inner brackets contain parts which are independent of $e_{1-4}$, and
can be calculated for given polarization vectors $e_{5,6}$ and remain
constant when summing over $e_{1-4}$. The next level of brackets contains
parts which are independent of $e_{1-2}$. Actually, on a DEC Station 5000
this matrix element can be evaluated roughly 1000 times per minute.
In fact, matrix element was calculated both in non-linear and
't~Hooft-Feynman gauges and these two results coincided with the machine
accuracy. In addition, electromagnetic gauge invariance of (\ref{ME}) was
checked numerically ${\cal M}(e_{1,2}\to p_{1,2})=0$.

\section{Cross sections}

Total cross sections for $W^+W^-ZZ$ and $W^+W^+W^-W^-$ production in
photon-photon collisions as a function of $\g\g$ center-of-mass energy for
different initial and final polarization states and two values of the Higgs
boson mass $m_H=100$~GeV and $m_H=\infty$ are shown in Figs.~6, 7. Different
combinations like $TTLL$ and $TLTL$ refer to a given order of $W^\pm$'s and
$Z$'s: {\it e.g.}, $TLTL$ denotes $W_T^\pm W_L^\mp Z_T Z_L$ or $W^\pm_T
W^\pm_L W^\mp_T W^\mp_L$ final states.

For the $\g W^+W^-$ vertex we use Thomson limit coupling $\alpha$, while all
the other couplings are derived from $\alpha(M_Z)$, $\sin\theta_W$ and
$\cos\theta_W$. So, we choose $\alpha^2\alpha(M_Z)^2$ as the overall
coupling factor.  Throughout this paper, we use the following set of
electroweak parameters:
\begin{eqnarray}
&\alpha = 1/137.036, & \alpha(M_Z) = 1/128.82, \nonumber\\
&M_W=80.22 \mbox{~GeV}, & M_Z=91.173\mbox{~GeV}, \nonumber\\
&\cos\theta_W = M_W/M_Z.&
\end{eqnarray}
Monte-Carlo numerical integration and event generation were performed by the
program package BASES/SPRING \cite{BASES}.

One can see that the cross sections are slightly larger for equal initial
photon helicities. The dominating contribution comes from four transverse
gauge boson production. The larger is the number of longitudinal gauge
bosons, the smaller is the cross section. However, longitudinal gauge bosons
do not decouple at large energies and all the cross sections rise with
energy even for light Higgs boson. For example, for $WWWW$ production the
ratio of $TTTT/TTTL$ is about $(60\div 70)$\%. The large fraction of
longitudinal polarization states production  was also observed earlier for
$\gamma\gamma\to WWZ$ reaction \cite{WWZ}.  The contributions from two
longitudinal weak bosons $TTLL$ and $TLTL$ are about an order of magnitude
smaller than that for $TTTT$ production. The yield of $TLLL$ and $LLLL$
final states is even smaller, however one can see that for infinite Higgs
boson mass their contribution can be orders of magnitude larger than for a
100~GeV Higgs. From Fig.~6 it is clear that large Higgs boson mass mainly
affects the $WWZZ$ cross sections with at least two longitudinal $Z$ bosons,
{\it e.g.}, for $m_H=\infty$ $TTLL$ cross section increases by a factor of
four at $\sqrt{s}=2$~TeV, while $TLTL$ and $LLTT$ remain almost the same.
For $WWWW$ production both $TLTL$ and $TTLL$ cross sections increase (see
Fig.~7), but at large energy the $TTLL$ cross section becomes dominating.
This is a consequence of the well known fact that for infinitely heavy Higgs
mass cross section of like charge $W^\pm_LW^\pm_L\to W^\pm_LW^\pm_L$
scattering is larger than opposite charge $W^+_LW^-_L\to W^+_LW^-_L$
scattering. Cross sections with transverse final gauge bosons, longitudinal
``spectator" $W$ bosons, $Z_TZ_T$, $Z_LZ_T$ or $W_TW_TW_TW_L$ final
states are practically the same for $m_H=100$~GeV and $m_H=\infty$.

It is this rise of the cross section of longitudinal electro-weak gauge
boson interactions that signals strong EWSB scenario \cite{SB}. This effect
is illustrated in Fig.~8 where production cross sections of $TTTT+TTTL$ as
well as of final states containing at least two longitudinal gauge bosons
are compared for $m_H=100$~GeV and $m_H=\infty$.  As usual, we can define
the heavy Higgs boson signal to be the difference between the cross section
with a heavy Higgs boson and the result with a light Higgs boson, {\it
e.g.},
\begin{equation}
\sigma(\mbox{signal for } m_H=\infty) =
\sigma(m_H=\infty)-\sigma(m_H=100\mbox{~GeV}).
\end{equation}
Consequently, cross section for light Higgs boson represents the background.
{}From Fig.~8 one can conclude that signal-to-background ratio is about 10\%
for total cross sections.

The relative ratios of $WWZZ$ and $WWWW$ cross sections can be qualitatively
understood using EWA. At high energy the following relations hold for $WW$,
$ZZ$ scattering cross sections integrated over $p_\perp>p_\perp^{min}>M_W$,
$M_Z$ for $m_H\sim M_W$:
\begin{equation}
\sigma(W^+ W^- \to W^+ W^-)
\simeq \sigma(W^\pm W^\pm \to W^\pm W^\pm)
\simeq \frac{1}{\cos^4\theta_W}\sigma(W^+ W^- \to Z Z),
\end{equation}
{\it i.e.} $WW\to ZZ$ cross section is about two times smaller that the
other cross sections. Therefore, the ratio of the cross sections of
transverse $WWWW$ and $WWZZ$ production is
\begin{eqnarray}
&&\left.
\frac{\sigma(\g\g\to W W W W)}{\sigma(\g\g\to W W Z Z)}
\right|_{m_H\sim M_W} \sim \\
&&\frac{\int dz_1dz_2 f_{W/\g}(z_1)f_{W/\g}(z_2)
\left[\sigma\left(W^+ W^- \to W^+ W^-\right)
+ \sigma\left(W^+ W^+ \to W^+ W^+\right)\right]}
{\int dz_1dz_2 f_{W/\g}(z_1)f_{W/\g}(z_2)
\sigma(W^+ W^- \to Z Z)}\sim 4. \nonumber
\label{WWWW/WWZZ}
\end{eqnarray}
On the other hand, ratio of the cross sections of $WWZZ$, $WWWW$ production
through Higgs resonance is given by
\begin{equation}
\left.
\frac{\sigma(\g\g\to W W W W)}{\sigma(\g\g\to W W Z Z)}
\right|_{resonance} =
\frac{\Gamma(H\to W^+W^-)}{\Gamma(H\to ZZ)}\sim 2.
\end{equation}
So, we conclude that signal-to-background ratio should be about two times
larger for $WWZZ$ production than for $WWWW$ production:
\begin{equation}
\left.\frac{S}{B}\right|_{WWZZ} \sim 2\,\left.\frac{S}{B}\right|_{WWWW}.
\label{HZZ/HWW}
\end{equation}

\section{Signal of heavy Higgs boson at photon linear collider}

Results presented in this section are obtained taking into account realistic
photon spectrum \cite{gg}. We assume that the product of electron and laser
photon helicity is $2\lambda_e\lambda_\g =2\lambda_e'\lambda_\g' = -0.9$ and
$\lambda_\g \lambda_\g' = 1$, so that final high energy photons are produced
predominantly in $J_Z=0$ states.

In Table~1 we summarize total cross sections for $WWWW$, $WWZZ$ production
at photon-photon collider realized at 1.5, 2 and 3~TeV linear collider.
Cross sections are quite large, for example more than ten thousand events of
four weak boson production will take place at 2~TeV linear collider with
$\int{\cal L}dt = 200$~fb$^{-1}$. As expected from (\ref{WWWW/WWZZ}), for
$m_H=100$~GeV the $WWWW$ cross section is about four times larger than
$WWZZ$ cross section.

The scattering reaction (\ref{WWWW}) leads to two scattered $W$'s or $Z$'s
emerging at large transverse momentum in the final state accompanied by two
``spectator" $W$'s at low $p_\perp$ focussed along the beam direction. We
assume that hadronic decay modes of the central pair will be observed, {\it
i.e.} we will not distinguish $W$'s from $Z$'s. The heavy Higgs signal can
be observed in the invariant mass spectrum of the two hard scattered weak
bosons. To select these $W$'s or $Z$'s we label all the final gauge bosons
according to their pseudo-rapidities $\eta_i$:
\begin{equation}
|\eta_1|\geq |\eta_2|\geq |\eta_3|\geq |\eta_4|.
\label{Ordering}
\end{equation}
We are interested in the mass spectrum of the ``central" pair $m(V_3V_4)$,
where $V$ denotes $W$ or $Z$. The important point to note is that in the
framework of EWA the initial $W_L$'s participating in the $W_LW_L$
scattering have a $1/(p_\perp^2+M_W^2)^2$ distribution with respect to
incoming photons from which they are produced. This is to be contrasted with
a $p_\perp^2/(p_\perp^2+M_W^2)^2$ distribution of the initiating $W_T$'s,
leading, {\it e.g.}, to $W_TW_T$ scattering. Analogous effect is known to
take place for $W$ distribution in quark -- (anti-) quark or $e^+e^-$
collisions \cite{SB,Gold}. The softer $p_\perp$ distribution in the
$W_LW_L$ case has a useful consequence: the spectator $W$'s tend to
emerge with smaller $p_\perp$ and correspondingly smaller rapidity for
$W_LW_L$ scattering than those associated with the background processes of
$W_TW_T$ or $W_TW_L$ scattering. Therefore, we will divide four final gauge
bosons in two pairs of forward (backward) $V_1V_2$ and central $V_3V_4$
according to the ordering (\ref{Ordering}) and will impose different cuts on
these pairs. We require that $|\eta_{3,4}|<1$ and, in addition, veto hard
forward (backward) $W$'s $|\eta_{1,2}|>1.5$ to enhance the
signal-to-background ratio.  Although, all that we need is the invariant
mass spectrum of $M(V_3V_4)$, to separate four gauge boson production from
$\gamma\gamma\to W^+W^-$ and $\gamma\gamma\to W^+W^-Z$ backgrounds we will
also tag forward (backward) spectators $V_{1,2}$ in the region outside the
dead cone along the beam direction $|\eta_{1\div 4}|<\eta_0$, where $\eta_0$
is determined by the acceptance of experimental installation.

The experimental signature is then given by four central hard jets from
$V_3V_4$ decay with a branching ratio of 49\% and jets or leptons in forward
and backward regions from the decay of spectator $W$'s.  We have not
modelled $W$, $Z$ decays, so cuts will be imposed on momenta of vector
bosons. $W$, $Z$ pairs can be selected using the good knowledge of the $W$,
$Z$ mass. We assume that the procedure of $W$ pair reconstruction at 500~GeV
$e^+e^-$ collider \cite{WW} can be applied. From four central jets two jet
combinations are selected which have masses closest as possible to the $W$,
$Z$ mass:
\begin{equation}
{\rm min}\left[(M_{(1)}-M_{W,Z})^2+(M_{(2)}-M_{W,Z})^2\right].
\end{equation}
The masses scatter significantly: only 38\% of the $W$, $Z$ pair events are
in the region $|M_{(1,2)}-M_{W,Z}|<10$~GeV \cite{WW}. We assume that the
same efficiency of central $W$, $Z$ mass reconstruction is applied for our
study. Reconstructing $W$, $Z$ masses most of the QCD backgrounds should be
rejected. Backgrounds from $\g\g\to t\bar t$ can be eliminated by top-quark
mass reconstruction and $b$-tagging \cite{WW}.

For the reaction $\g\g\to WWZZ$ it is possible to test our procedure of
separation of hard scattered $Z$'s from spectator $W$'s based on
(\ref{Ordering}). In Figs.~9-11 various distributions of $W$, $Z$ bosons and
corresponding distributions based on labeling (\ref{Ordering}) are shown at
$\sqrt{s_{e^+e^-}}=2$~TeV and $m_H=1$~TeV and 100~GeV. It is assumed that
either $M(ZZ)$ or $M(V_3V_4)$ is greater that 500~GeV.  In Fig.~9
pseudo-rapidity distributions are presented for $Z$ and $W$ bosons as well
as corresponding distributions for mean rapidity of $V_3V_4$ and $V_1V_2$
pair. $Z$ bosons have central rapidity distribution peaking at $\eta_Z=0$.
Spectator $W$'s have rapidity distribution peaking in the forward-backward
direction. For $m_H=1$~TeV the $W$ distribution peaks at $\eta_W\simeq\pm
1.5$, while for 100~GeV Higgs the background peaks at $\eta_W\simeq\pm 1$.
The $\langle\eta_{3,4}\rangle$ and $\langle\eta_{1,2}\rangle$ distributions
roughly follow $\eta_Z$ and $\eta_W$ distributions, respectively. Fig.~10
presents $p_\perp$ distributions. $p_\perp^W$ and $p_\perp^{1,2}$
distributions are peaking at small values of $p_\perp<100$~GeV, while
$p_\perp^Z$ and $p_\perp^{3,4}$ distributions are wide and peak at
$p_\perp\sim 250$~GeV.  The 1~TeV Higgs signal enriches large- (low-)
$p_\perp$ tail of the $p_\perp^Z$ ($p_\perp^W$) distributions. From Fig.~11
one can see that the transverse momentum distributions of $ZZ$ and $V_3V_4$
pair are again quite similar. Unlike the $p_\perp^Z$ ($p_\perp^{3,4}$)
distributions, most of the 1~TeV Higgs signal lies at $p_\perp^{ZZ}$,
$p_\perp^{3+4}<200$~GeV.  Figs.~12-13 show the pseudo-rapidity and transverse
momentum distributions for the reaction $\g\g\to WWWW$ for
$M(V_3V_4)>500$~GeV. They look similar to corresponding distributions in
Figs.~9-11.

The $M(V_3V_4)$ invariant mass distribution for the signal ($m_H=1$~TeV) and
background ($m_H=100$~GeV) are shown in Fig.~14 at $\sqrt{s_{e^+e^-}}=2$~TeV
assuming that the annual integrated luminosity is 300~fb$^{-1}$. The
luminosity is derived from rescaling of 20~fb$^{-1}$ for 500~GeV NLC to keep
a roughly constant event rate for $\sigma_{\rm point}=4\pi\alpha^2/3s$. The
rapidity cuts $|\eta_{1\div 4}|<3$, $|\eta_{3,4}|<1$ and $|\eta_{1,2}|>1.5$
are imposed. The enhancement for $m_H=1$~TeV in the region
$M(V_3V_4)>500$~GeV is clearly seen in all the histograms. As predicted from
(\ref{HZZ/HWW}), the signal-to-background ratio is about two times larger
for $WWZZ$ than for $WWWW$ reaction. Also, for $\g\g\to WWZZ$ reaction the
$M(V_3V_4)$ invariant mass spectrum looks almost the same as $M(ZZ)$ one.

We summarize our results for $m_H=1$~TeV and $m_H=\infty$ in Table 2. This
table gives the signal and background event rates as a function of the dead
cone along the photon beams direction. The pseudo-rapiditiy cuts $\eta_0=3$,
2.5 correspond to approximately $5^\circ$ and $10^\circ$ dead cone,
respectively. Hadronic branching ratio of 49\% and 38\% efficiency of the
$W$, $Z$ pair mass reconstruction are included. From the Table~2 is apparent
that a 1~TeV Higgs boson can be observable at 2~TeV PLC at 5$\sigma$ level
even for large $10^\circ$ dead cone.  Comparing Tables~1 and 2 one can see
that while the signal contributes only about 10\% to the total cross section
for $m_H=1$~TeV, appropriate cuts permit to enhance the signal-to-background
ratio by an order of magnitude.  While the signal-to-backgroud ratio is two
times larger for $WWZZ$ final state, due to a four times larger statistics
reaction $\g\g\to WWWW$ gives almost the same statistical significance of
the Higgs signal.  The ``spectator" $W$ veto $\eta_{1,2}>1.5$ enhances the
$S/B$ ratio in 2-3 times, but at the expense of large loss in statistics, so
that statistical significance is almost the same with or without the forward
$W$ veto. It is important to have as good as possible coverage in
forward-backward directions. Changing $\eta_0$ from 3 to 2.5 we diminish the
signal in 1.5-2 times. As for infinitely heavy Higgs boson, the statistical
significance of the signal is always below 3$\sigma$ for any realistic
detector acceptance.

Fig.~15 and Table~3 show the signal and background for $m_H=500$~GeV and
$m_H=700$~GeV at $\sqrt{s_{e^+e^-}}=1.5$~TeV assuming the annual integrated
luminosity of 200~fb$^{-1}$. The pronounced peak from 500~GeV Higgs boson
should be easily observable with statistical significance greater than
10$\sigma$. For $m_H=700$~GeV the peak is already quite wide, but still it
should be observable at the level of about 4$\sigma$. So, in principle, the
reactions of $WWWW$, $WWZZ$ production in photon-photon collisions allow to
observe Higgs boson heavier than 400~GeV, which is the maximum Higgs mass
detectable in the reaction $\g\g\to ZZ$ \cite{aazz}. However, for the former
case Higgs boson production emerges from the $W$ fusion reaction and has
nothing to do with two-photon Higgs width, which can be measured for lighter
Higgs in $\g\g\to ZZ$ reaction.  It is hardly possible to push the
observable Higgs mass well above 700~GeV at 1.5~TeV machine.

Finally, to exemplify PLC potential to probe nonresonant strong $W_LW_L$,
$Z_LZ_L$ scattering in Fig.~16 and Table~4 we show results for infinitely
heavy Higgs boson at $\sqrt{s_{e^+e^-}}=3$~TeV assuming the annual
integrated luminosity of 300~fb$^{-1}$. To avoid problems with non-unitarity
of cross section of the longitudinal gauge boson scattering at high energy
we assumed that $M(V_3V_4)<1.5$~TeV.  The excess of events for $m_H=\infty$
should be observable at 4$\sigma$ level even for the worst detector
coverage.

\section{Conclusions}

We have demonstrated that significant signal from 1~TeV Higgs resonance can
be observed in the hadronic final states in photon-photon collisions at
2~TeV linear collider for integrated luminosity of 300~fb$^{-1}$. The
nonresonant strong scattering of longitudinal weak gauge bosons can be
studied at larger collision energy of 3~TeV. Higgs mass range up to 700~GeV
can be covered at 1.5~TeV PLC for integrated luminosity of 200~fb$^{-1}$.

The most important question is, certainly, comparison of the potential of
photon-photon collider with that of other machines. For hadronic and
$e^+e^-$ colliders much more detailed investigations were done including
decays of final $W$'s and $Z$'s and detector simulations
\cite{SB,Gold,kurihara}. For example, conclusion is done \cite{kurihara} that
the signal from 1~TeV Higgs boson is distinguishable from the case of
massless Higgs at the center-of-mass energy of $e^+e^-$ collider of 1.5~TeV
and integrated luminosity of 200~fb$^{-1}$. However, it is also found that
the integrated luminosity of 310~fb$^{-1}$ and 80~fb$^{-1}$ is needed to
discriminate the $m_H=\infty$ signal at $3\sigma$ level at 2~TeV and 3~TeV
linear collider, respectively.  So, we can very roughly estimate that
potential of 2~TeV linear collider in photon-photon mode is at least the
same as that of 1.5~TeV $e^+e^-$ collider, provided that their luminosities
are the same.  The optimistic conclusion of \cite{cheung} that luminosity of
10~fb$^{-1}$ could suffice to study strong EWSB in photon-photon collisions
at 2.5~TeV $e^+e^-$ collider is applicable only to resonant models of the
SB, {\it e.g.} SM Higgs boson with a mass of 1~TeV. To observe non-resonant
strong $W_L$, $Z_L$ scattering ($m_H=\infty$) in $\g\g$ collisions, which is
an adequate goal at $e^+e^-$ energy above 2~TeV, definitely the luminosity
above 100~fb$^{-1}$ is needed.  In addition, we should mention that
conclusions of \cite{SB,hagiwara,kurihara} are based on the assumption that
the background from $e^+e^-\to e^\pm\nu W^\mp Z$, which is comparable to the
signal, can be suppressed by distinguishing the $W$'s from $Z$'s in the
final state. But the accuracy of calorimetric measurement of the di-jet
invariant mass is as large as the intrinsic $W$--$Z$ mass difference and it
is not clear that high $W$--$Z$ separation efficiency can be achieved.

Finally, we would like to point out that if due to specific conditions at
the interaction point a huge luminosity $10^{35-36}$~cm$^{-2}$~s$^{-1}$ is
technically achievable in high energy photon-photon collisions
\cite{gg1,gg2,gg3} and if it will be possible to make experiments at such a
luminosity, photon-photon option will become very competitive with normal
$e^+e^-$ mode of linear collider.

\section*{Acknowledgements}

I am grateful to M.~Berger, E.~Boos, F.~Boudjema, S.~Brodsky,
M.~Chanowitz, K.~Cheung, I.~Ginzburg, T.~Han, V.~Ilyin, F.~Richard,
V.~Serbo and V.~Telnov for many helpful discussions. This work was
supported, in part, by the INTAS-93-1180 grant.

\newpage
\section*{Figure captions}
\parindent=0pt
\parskip=\baselineskip

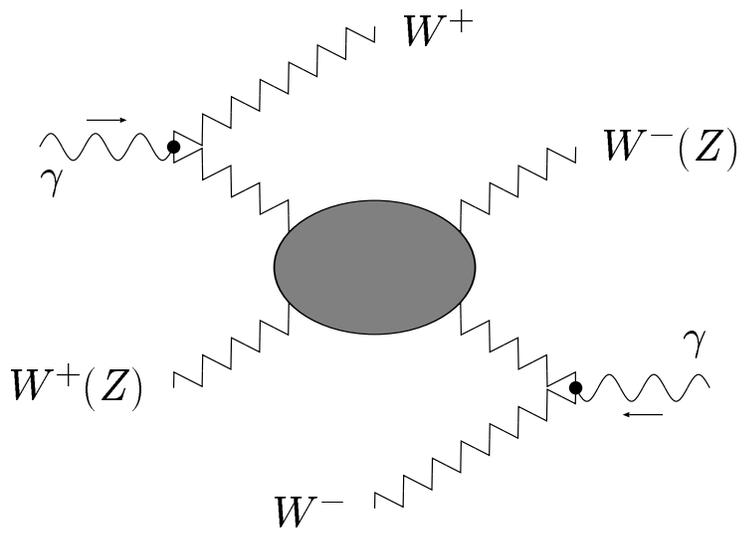
Fig.~1. $WW$ scattering at a photon-photon collider.

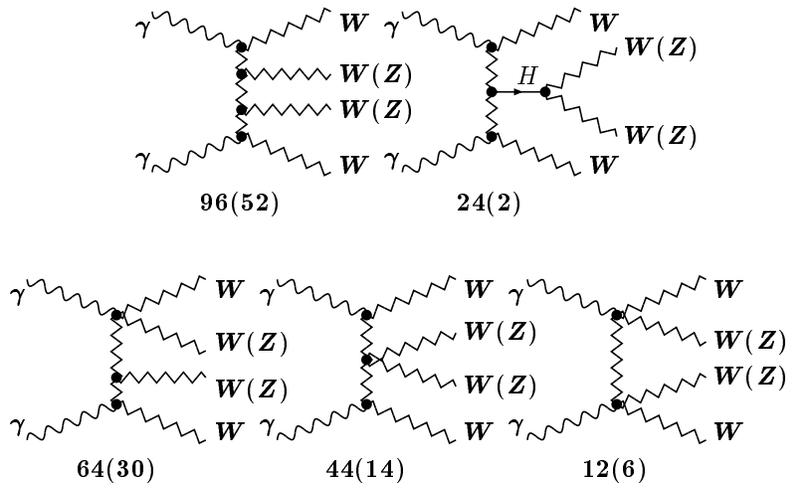
Fig.~2. Typical Feynman diagrams corresponding to five different topologies
contributing to the reactions $\g\g\to WWWW$, $WWZZ$. The numbers below the
diagram denote the total number of graphs belonging to a given topology. The
numbers in parentheses refer to the reaction $\g\g\to WWZZ$

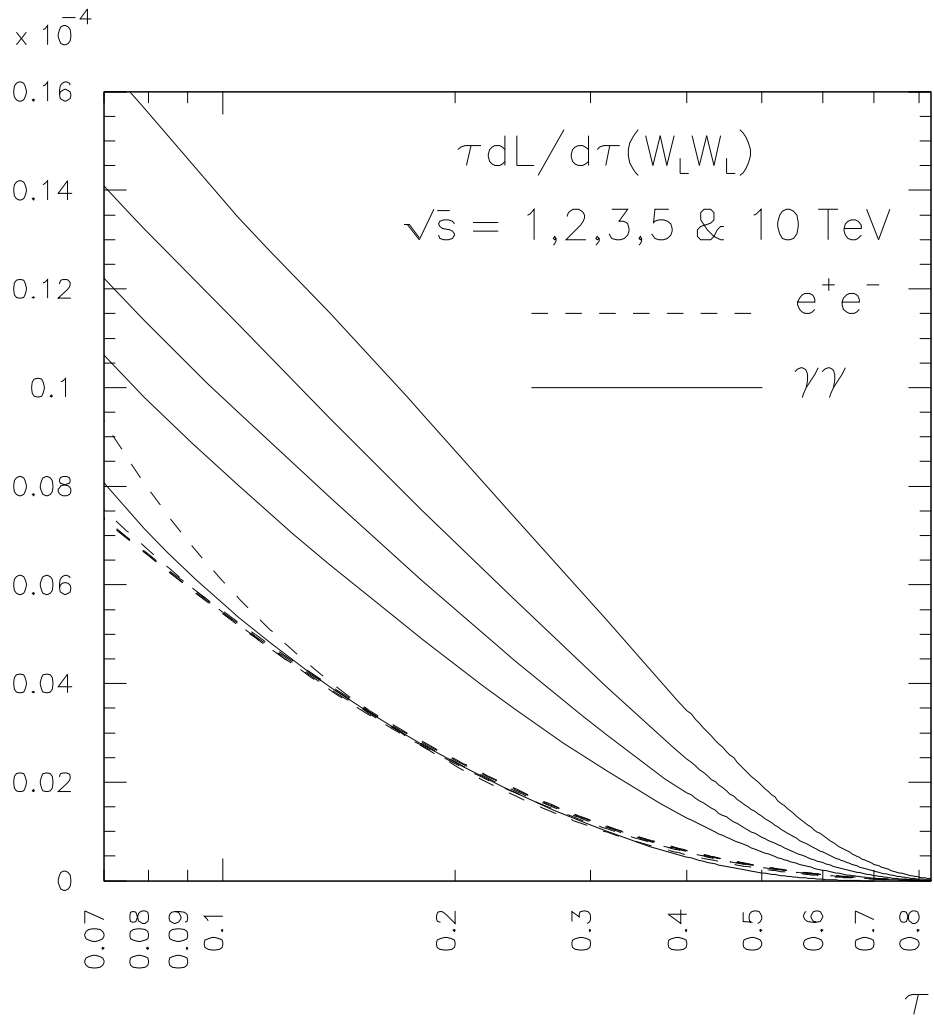
Fig.~3. The $W_LW_L$ luminosity in photon-photon and $e^+e^-$ collisions as
a function of $\tau=M_{W_LW_L}^2/s$ at the center-of-mass energy of 1, 2, 3,
5 and 10~TeV. For $e^+e^-$ case upper curves correspond to lower energy,
while for photon-photon case, on the contrary, upper curves correspond to
larger energy.

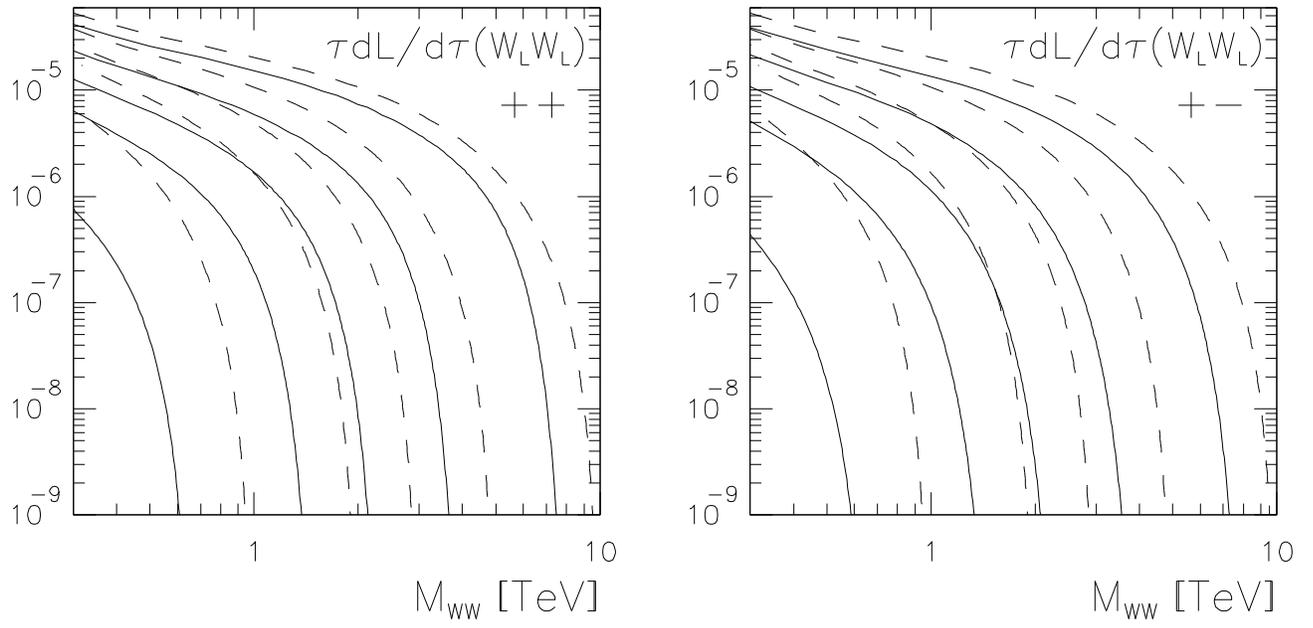
Fig.~4. The $W_LW_L$ luminosity in photon-photon collisions, calculated
taking into account realistic photon spectrum, and in $e^+e^-$ collisions as
a function of $M_{W_LW_L}$ at the center-of-mass energy of 1, 2, 3, 5 and
10~TeV. Upper curves correspond to larger energy. $++$ and $+-$ refer to
initial laser photon helicities $\lambda_\g\lambda_\g'=+1$ and $-1$,
respectively.

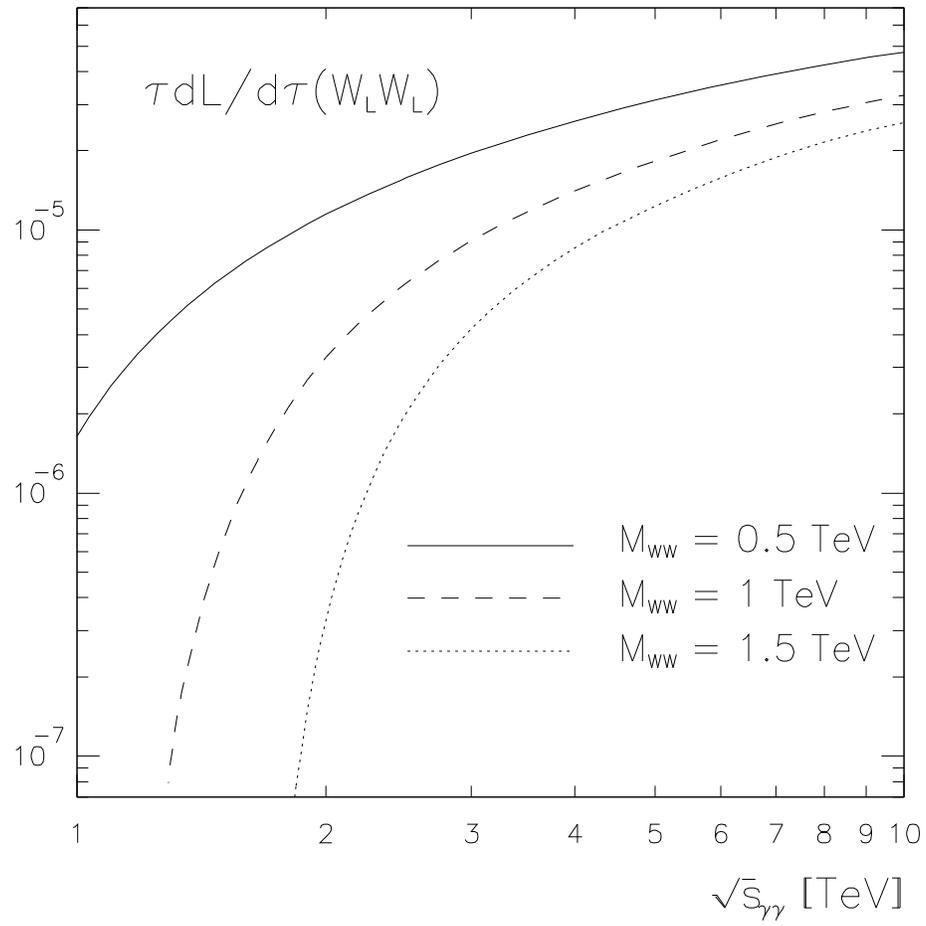
Fig.~5. The $W_LW_L$ luminosity in photon-photon collisions as
a function of $\sqrt{s_{\g\g}}$ for $M_{W_LW_L}=0.5$, 1 and 1.5~TeV.

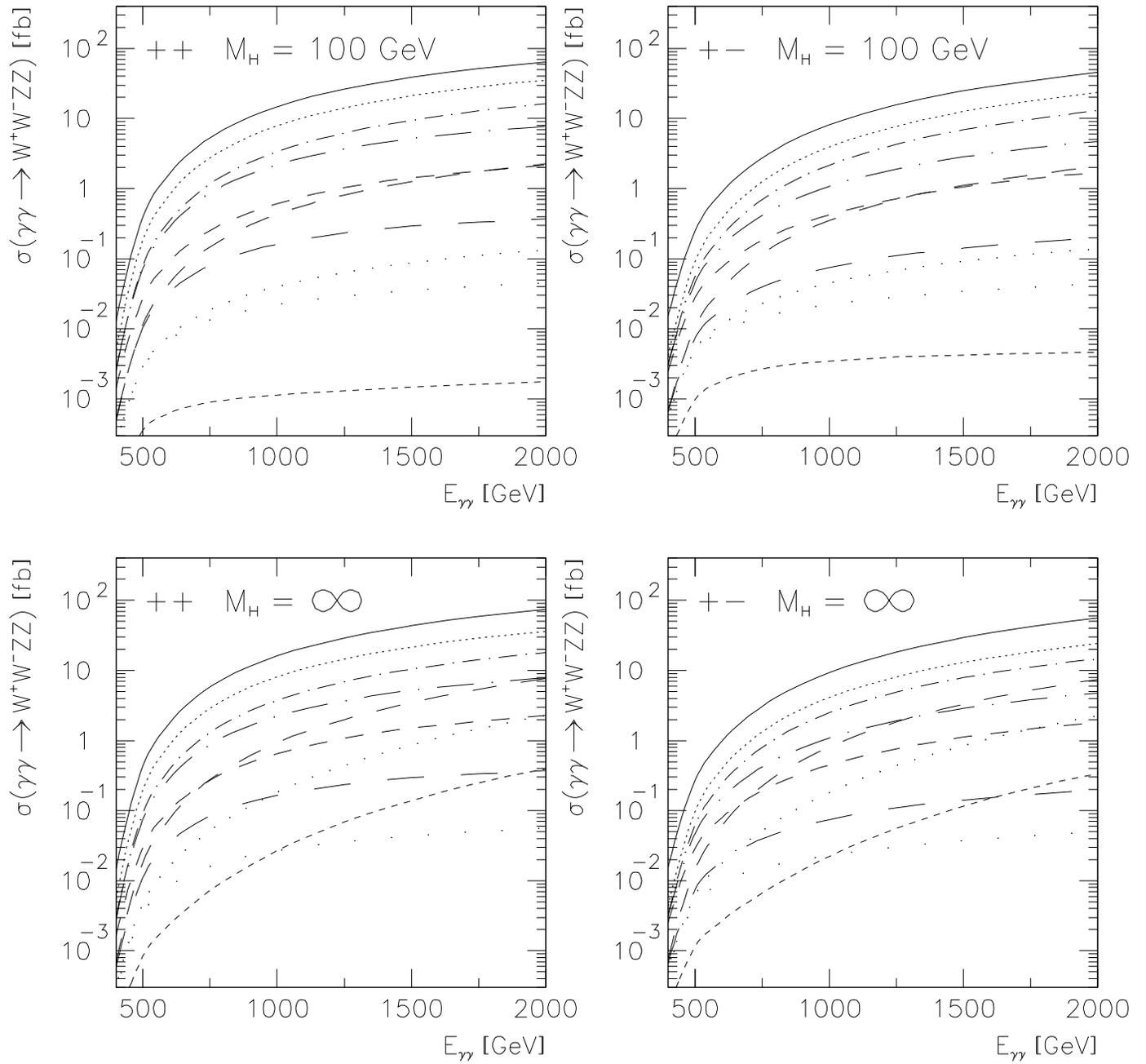
Fig.~6. Cross sections for different polarization states of initial
and final particles of the reaction $\gamma\gamma\to W^+W^-ZZ$ for
$m_H=100$~GeV and $m_H=\infty$ as a function of $\gamma\gamma$
center-of-mass energy. $++$ and $+-$ refer to initial photon helicities.
Solid line denotes total cross section. Also are shown curves for different
polarization states of $W^+W^-ZZ$:
$TTTT$~($\cdot$$\cdot$$\cdot$$\cdot$$\cdot$$\cdot$$\cdot$);
$TTTL$~($-\cdot -\cdot -\cdot$)
and $TLTT$~(--- $\cdot$ --- $\cdot$ ---);
$TLTL$~(-- -- -- --),
$TTLL$~($-\ -\ -$)
and $LLTT$~(---~~~---~~~---);
$TLLL$~($\cdot\cdot\cdot\cdot\cdot$)
and $LLLT$~($\cdot\ \cdot\ \cdot\ \cdot$);
$LLLL$~(- - - -).

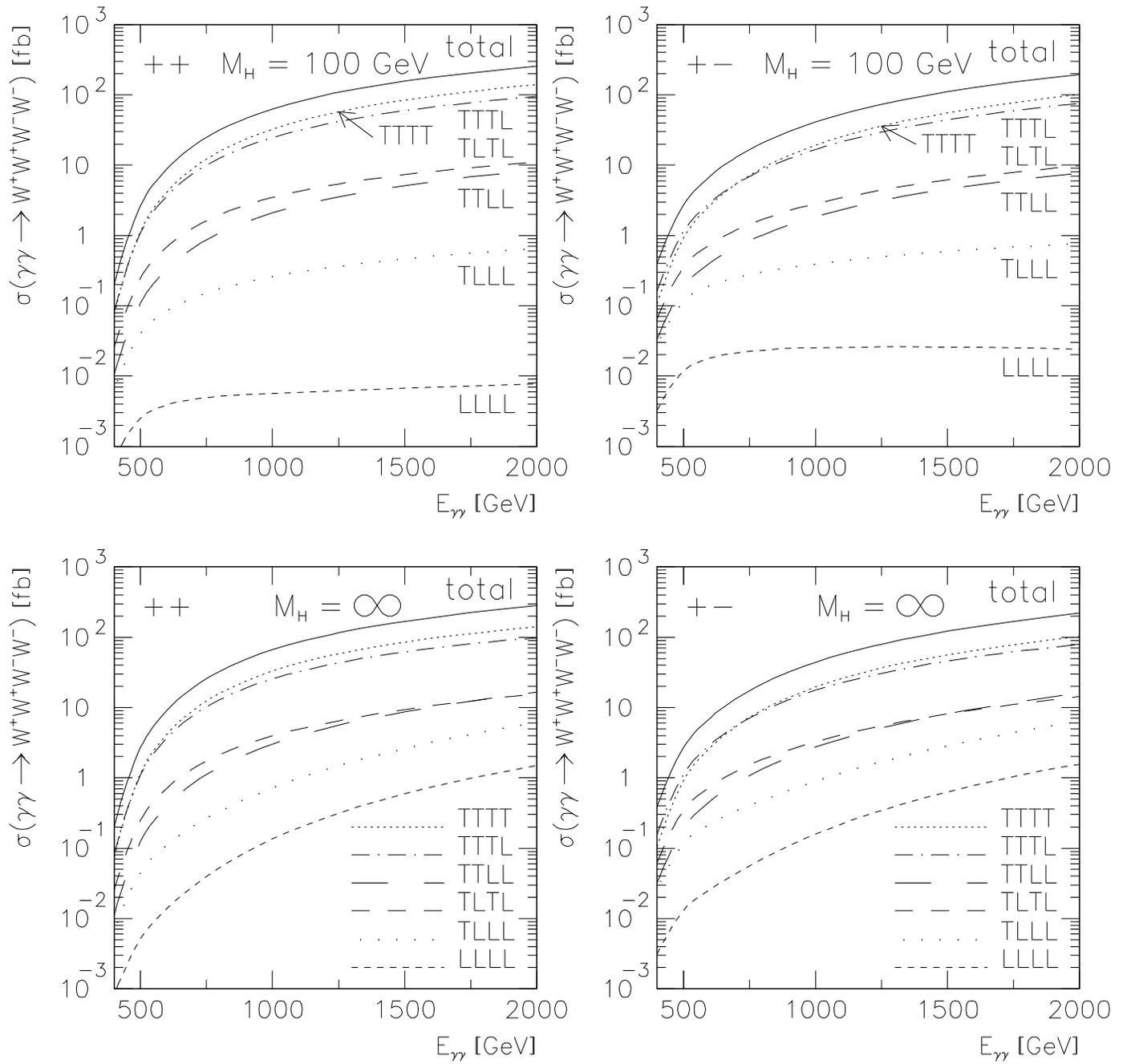
Fig.~7. Cross sections for different polarization states of initial
and final particles of the reaction $\gamma\gamma\to W^+W^+W^-W^-$ for
$m_H=100$~GeV and $m_H=\infty$ as a function of $\gamma\gamma$
center-of-mass energy.

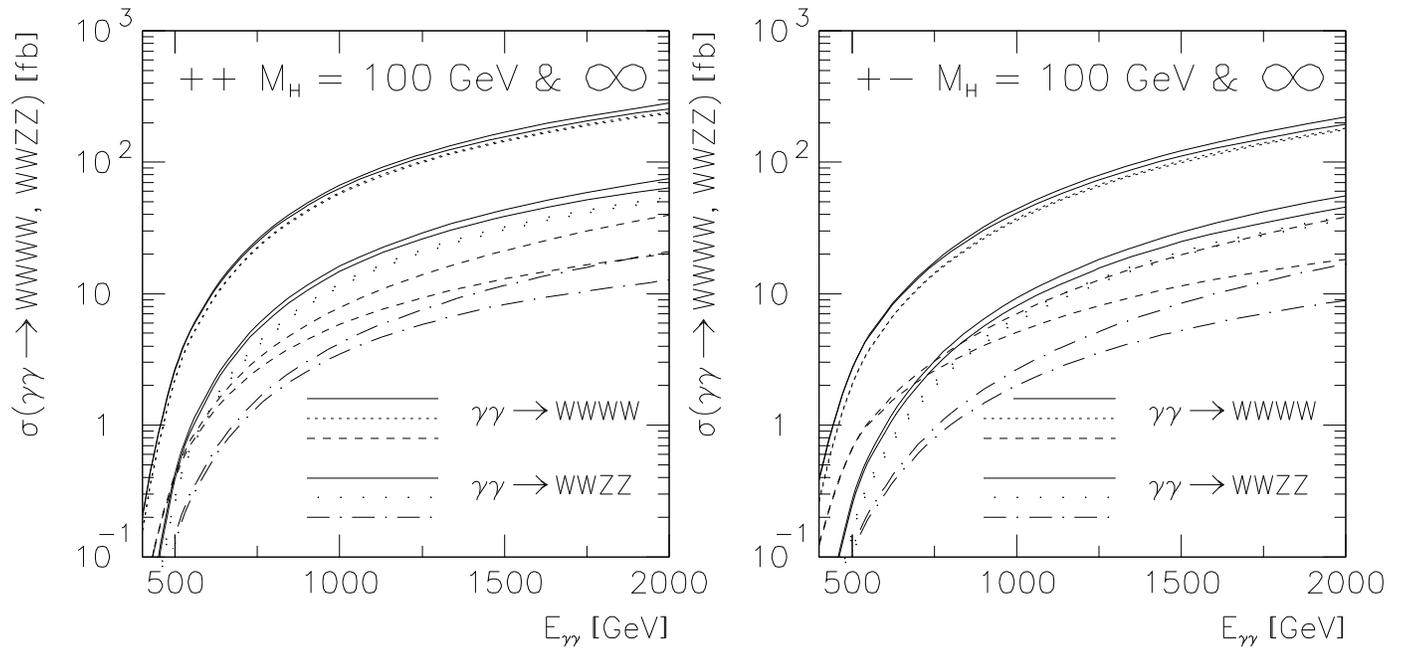
Fig.~8. Comparison between the cross sections for $m_H=100$~GeV and
$m_H=\infty$ for equal and opposite helicities of the initial photons. For
the reaction $\gamma\gamma\to WWWW$ the following cross sections are shown:
total cross section (solid line); the $TTTT+TTTL$ cross section (dotted
line); the sum of cross sections with at least two longitudinal final $W$'s
(dashed line). For the reaction $\gamma\gamma\to WWZZ$ corresponding cross
sections are denoted by solid, dotted and dash-dotted lines.

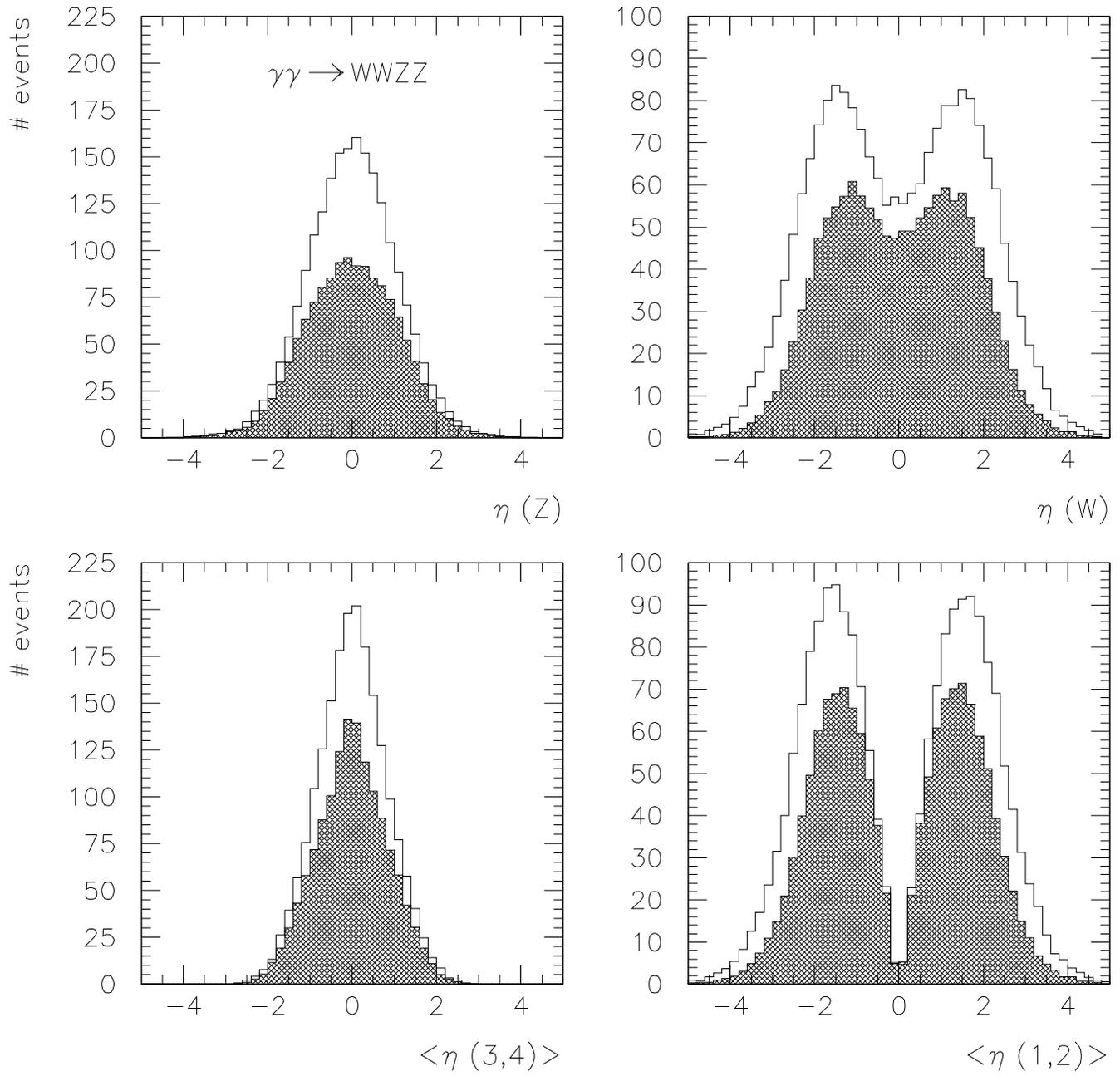
Fig.~9. Pseudo-rapidity distributions for the reaction $\g\g\to WWZZ$ for
$m_H=100$ (shaded histogram) and 1000~GeV at $\sqrt{s_{e^+e^-}}=2$~TeV for
integrated luminosity of 300~fb$^{-1}$.

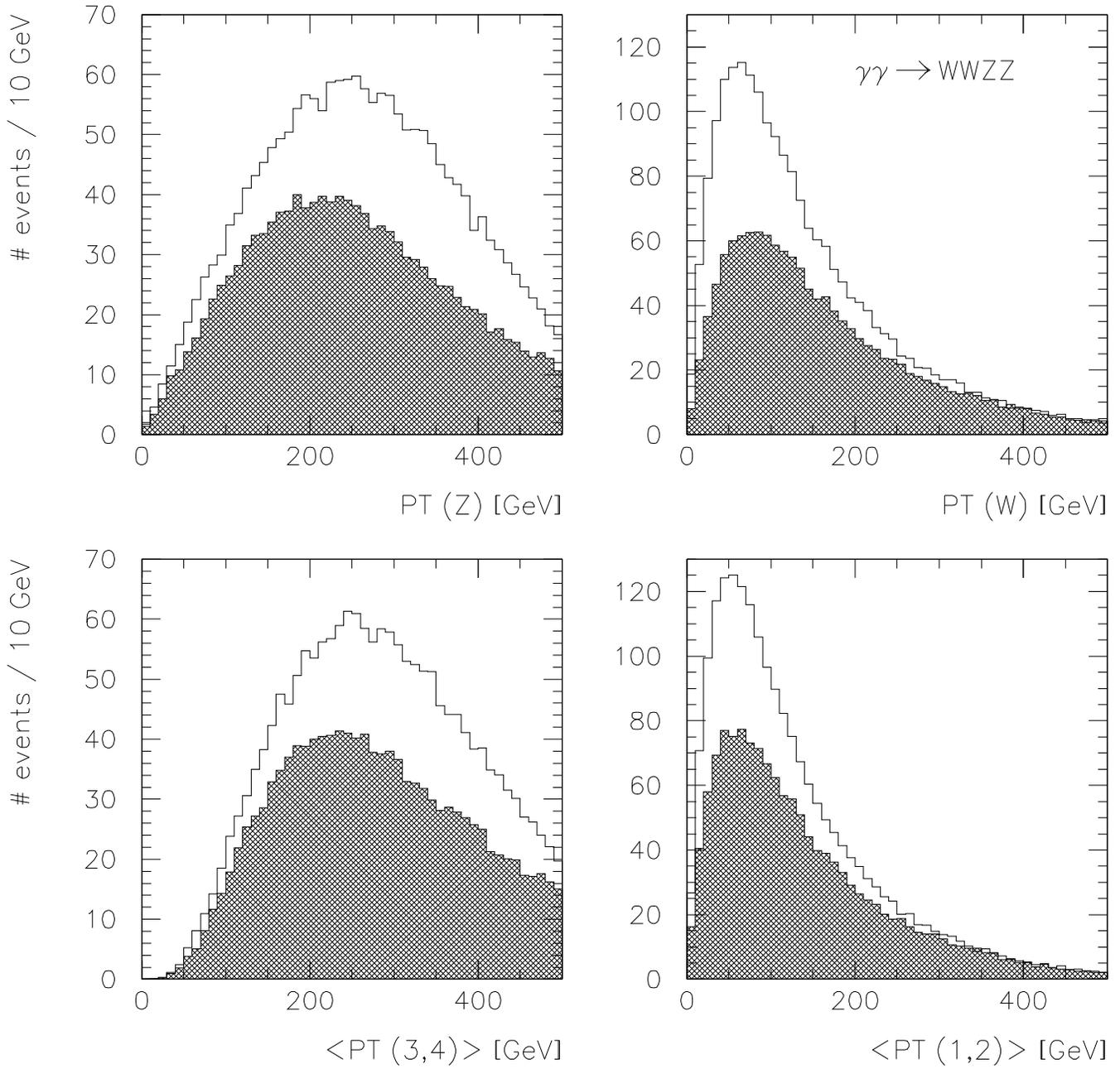
Fig.~10. $p_\perp$ distributions for the reaction $\g\g\to WWZZ$ for
$m_H=100$ (shaded histogram) and 1000~GeV at $\sqrt{s_{e^+e^-}}=2$~TeV for
integrated luminosity of 300~fb$^{-1}$.

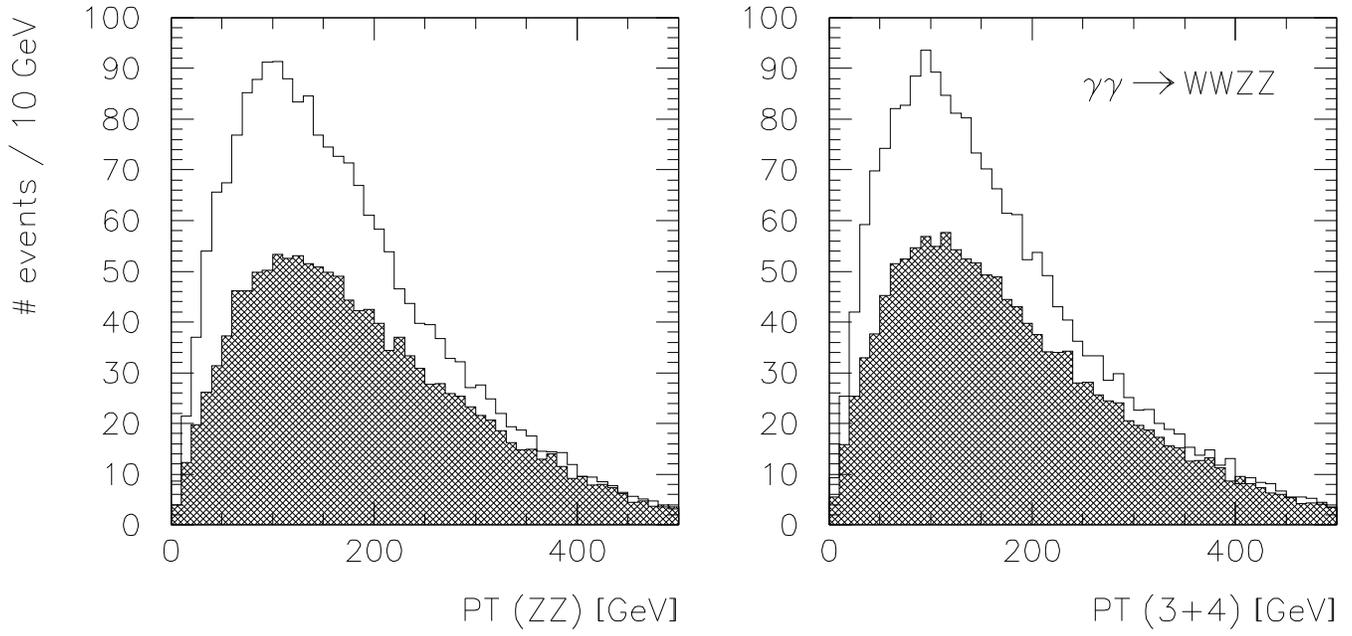
Fig.~11. $p_\perp^{ZZ}$ and $p_\perp^{3+4}$ distributions for the reaction
$\g\g\to WWZZ$ for $m_H=100$ (shaded histogram) and 1000~GeV at
$\sqrt{s_{e^+e^-}}=2$~TeV for integrated luminosity of 300~fb$^{-1}$.

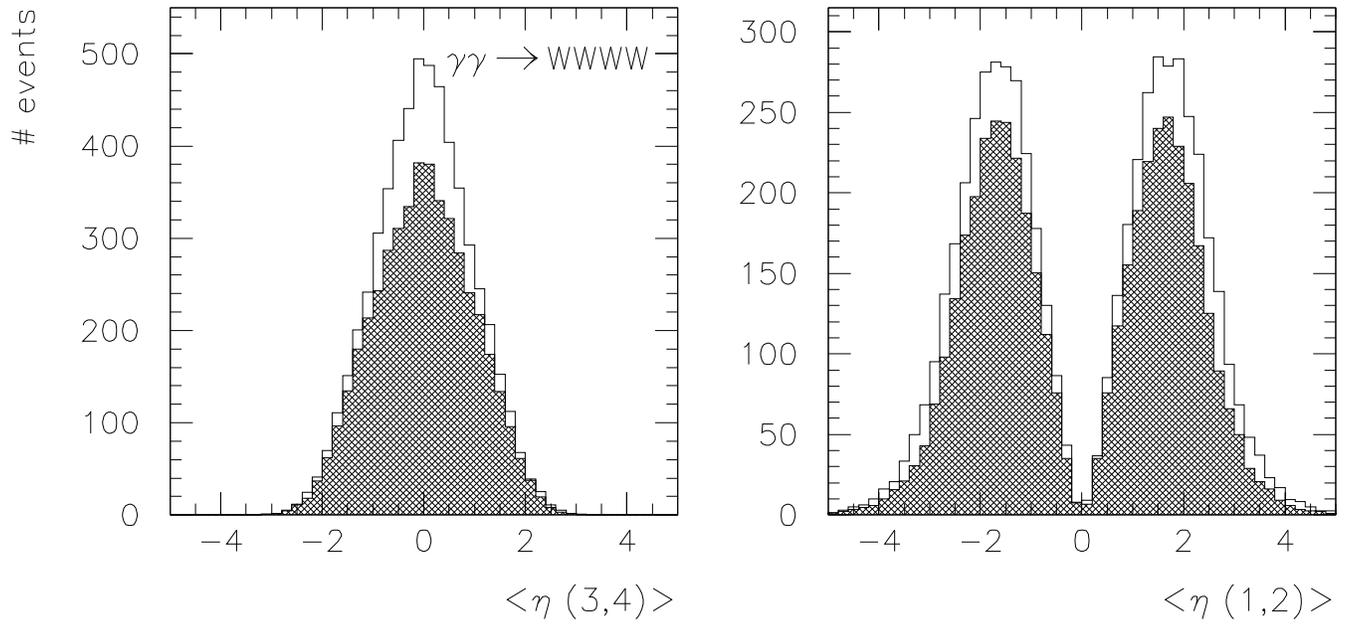
Fig.~12. Pseudo-rapidity distributions for the reaction $\g\g\to WWWW$ for
$m_H=100$ (shaded histogram) and 1000~GeV at $\sqrt{s_{e^+e^-}}=2$~TeV for
integrated luminosity of 300~fb$^{-1}$.

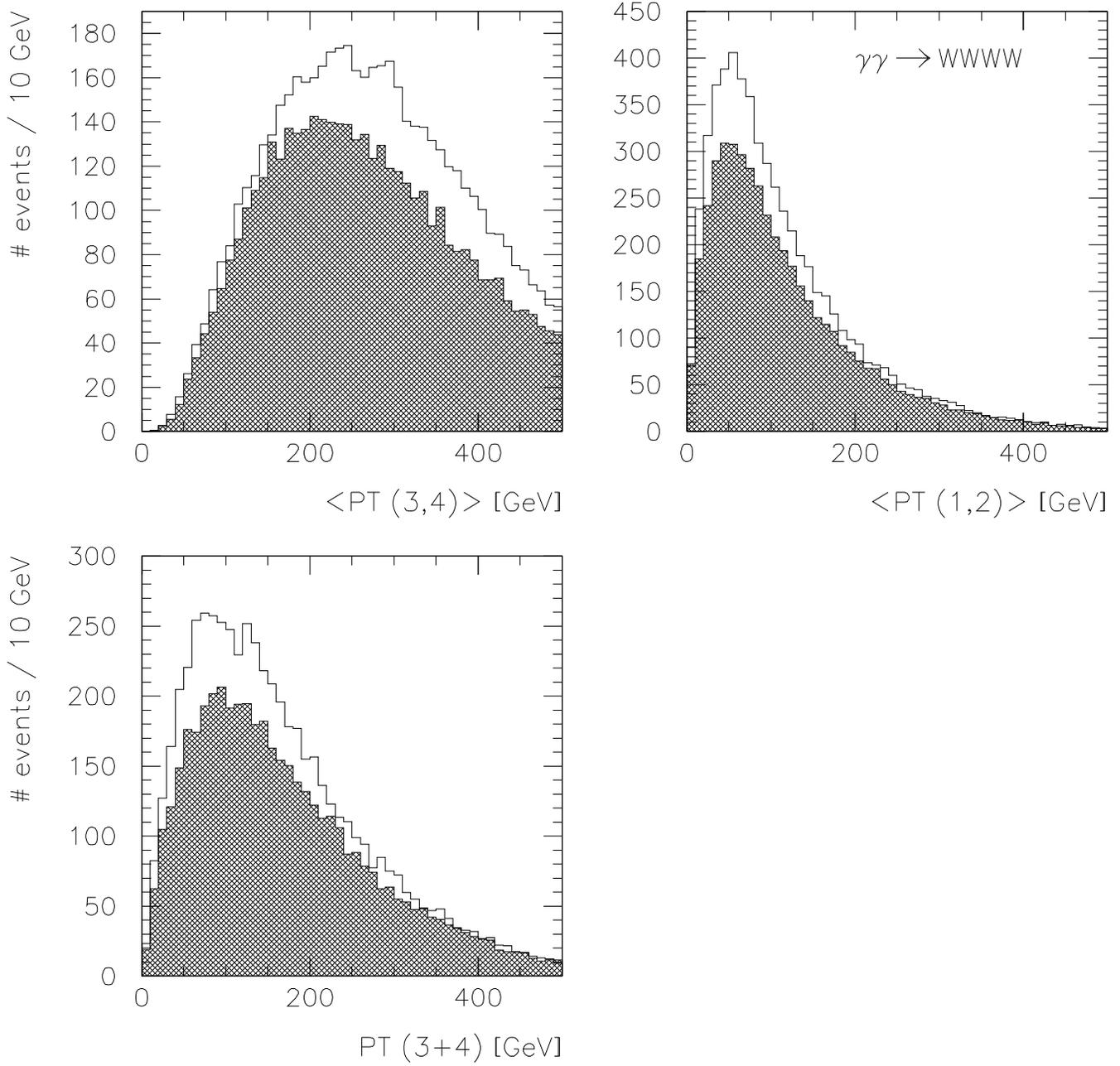
Fig.~13. $p_\perp$ distributions for the reaction $\g\g\to WWWW$ for
$m_H=100$ (shaded histogram) and 1000~GeV at $\sqrt{s_{e^+e^-}}=2$~TeV for
integrated luminosity of 300~fb$^{-1}$.

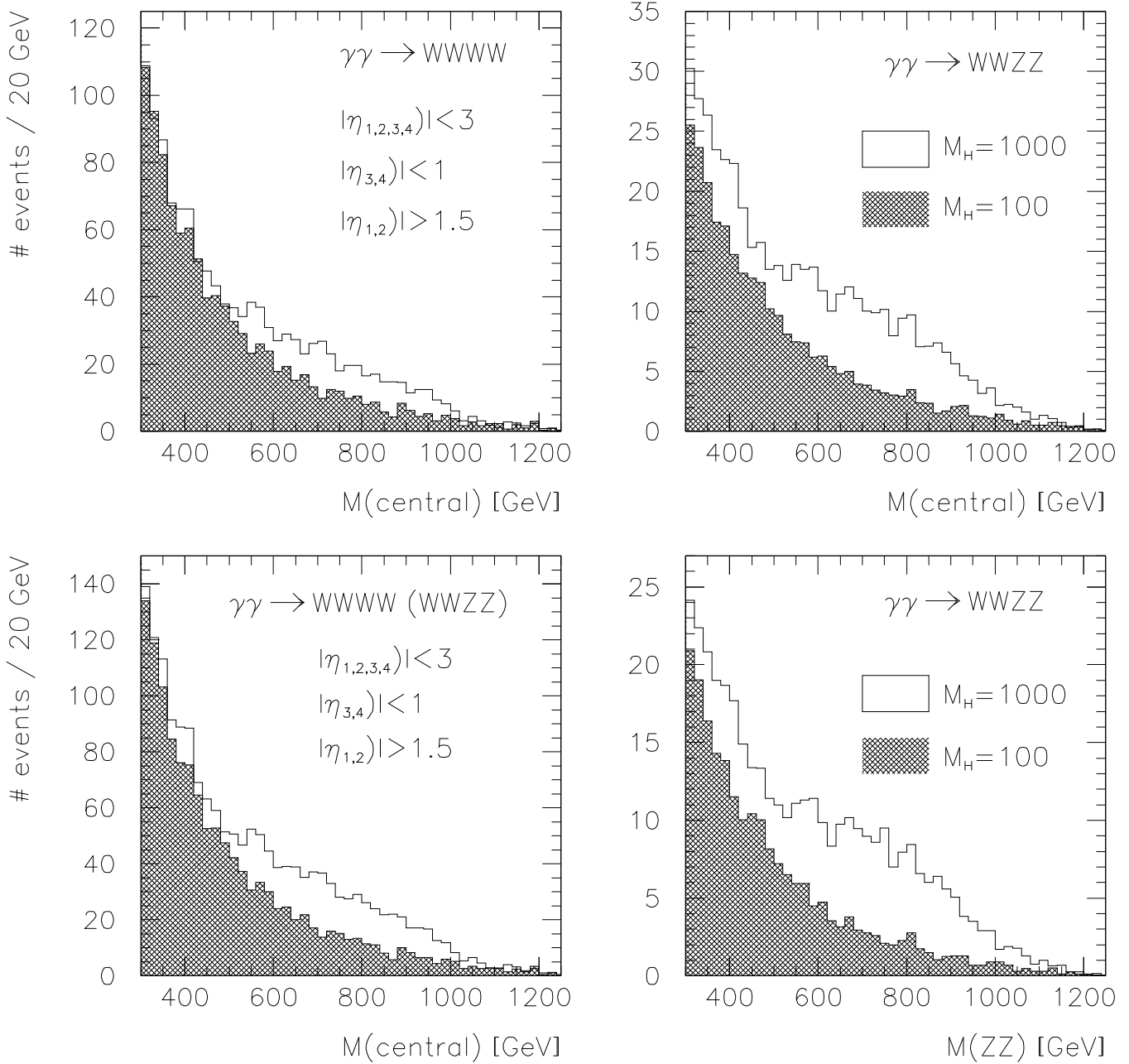
Fig.~14. Invariant mass $M(V_3V_4)$ distributions for $WWWW$, $WWZZ$ and
$WWWW+WWZZ$ production in $\gamma\gamma$ collisions at 2~TeV linear collider
for $m_H=100$ (shaded histogram) and 1000~GeV for integrated luminosity of
300~fb$^{-1}$.  The last histogram assumes that $W$'s are distinguished from
$Z$'s for $WWZZ$ production, the corresponding cuts are $|\eta_{1\div
4}|<3$, $|\eta_Z|<1$ and $|\eta_W|>1.5$. No branching ratios or efficiencies
are included.

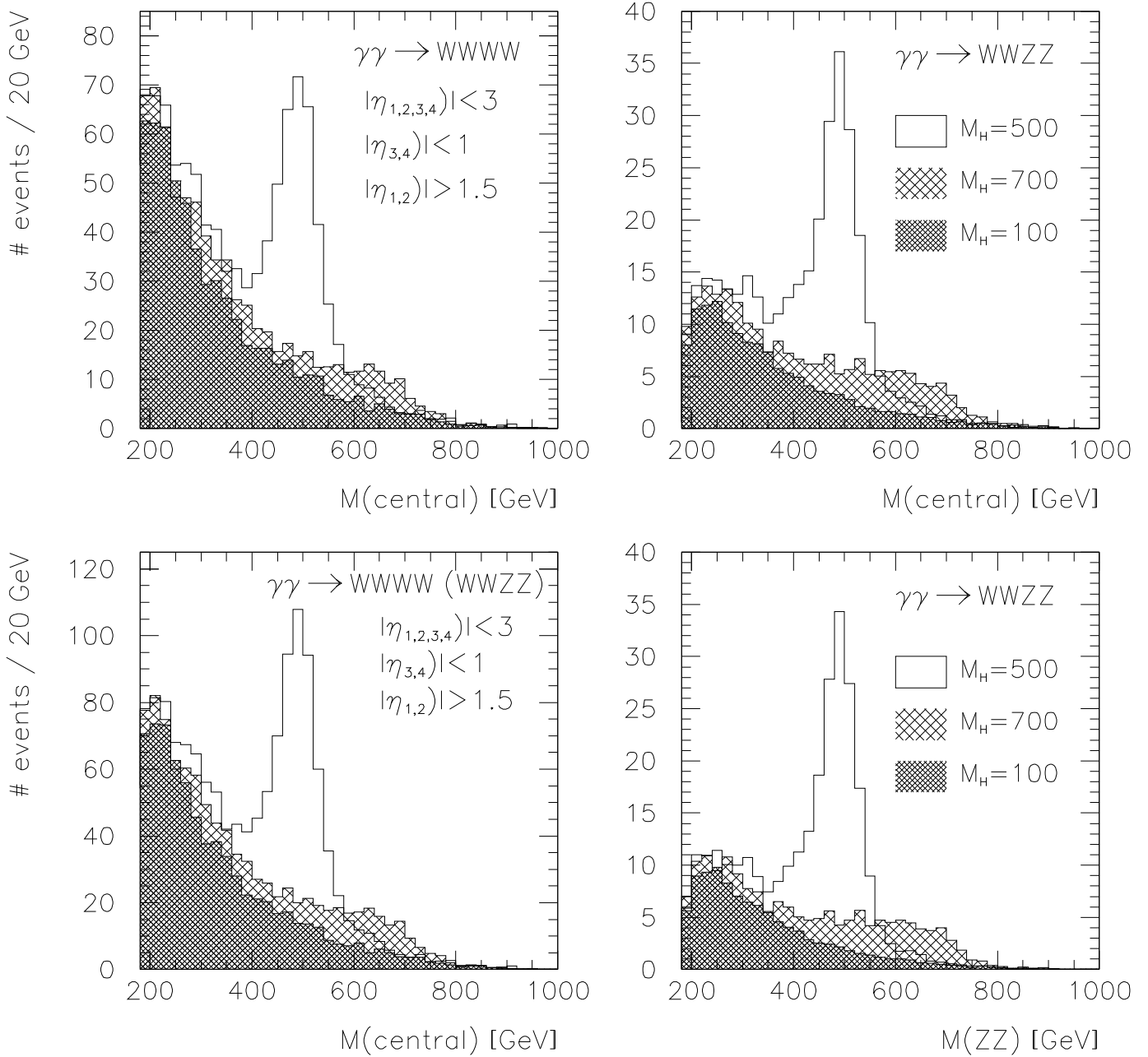
Fig.~15. Invariant mass distributions for $WWWW$, $WWZZ$ and $WWWW+WWZZ$
production in $\gamma\gamma$ collisions at 1.5~TeV linear collider for
$m_H=100$, 500 and 700~GeV for integrated luminosity of 200~fb$^{-1}$.

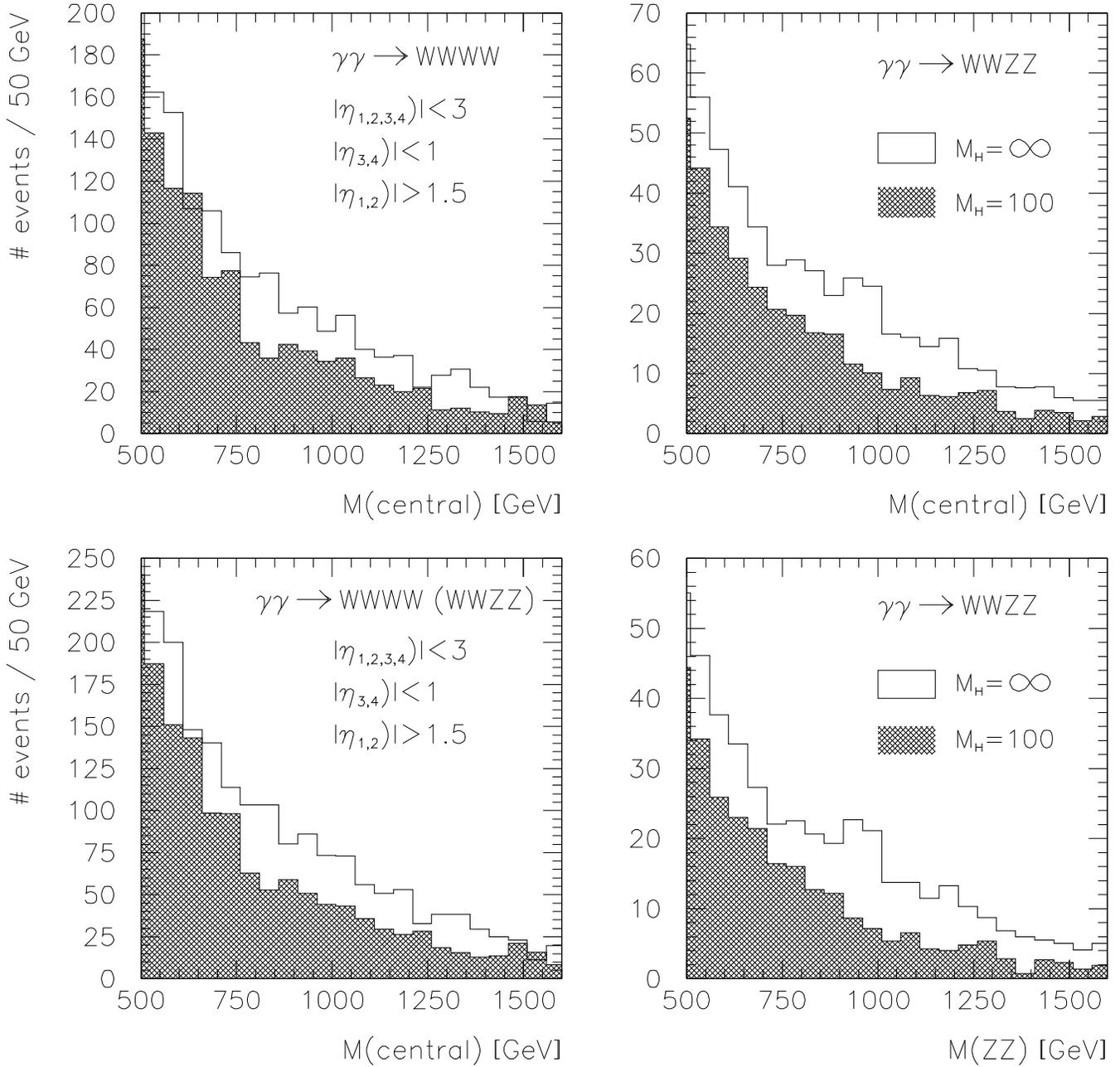
Fig.~16. Invariant mass distributions for $WWWW$, $WWZZ$ and $WWWW+WWZZ$
production in $\gamma\gamma$ collisions at 3~TeV linear collider for
$m_H=100$~GeV and $\infty$ for integrated luminosity of 300~fb$^{-1}$.

\newpage
\section*{Table captions}
\parindent=0pt
\parskip=\baselineskip

Table 1: Total cross sections (in fb)  for $\gamma\gamma\to W^+W^+W^-W^-$
and $\gamma\gamma\to W^+W^-ZZ$ reactions at $\sqrt{s_{e^+e^-}}=1.5$, 2 and
3~TeV.

Table 2: Event rates for signal ($S$) and background ($B$),
signal-to-background ratio and the number of standard deviations  for
$WWWW$, $WWZZ$ final states and their sum  at $\sqrt{s_{e^+e^-}}=2$~TeV,
$m_H=1$~TeV and $\infty$ for various values of the dead cone angle and
various cuts. The invariant mass $M_{34}$ of central pair is required to be
greater than 500~GeV. Branching ratio of 49\% for hadronic decays of central
$WW$, $ZZ$ pair and efficiency of central $W$, $Z$ mass reconstruction of
38\% are included.

Table 3: Events rates at $\sqrt{s_{e^+e^-}}=1.5$~TeV and $m_H=500$ and
700~GeV. For $m_H=500$~GeV the invariant mass $M_{34}$ of central pair is
required to be 400~GeV$<M_{34}<$600~GeV; for $m_H=700$~GeV
500~GeV$<M_{34}<$900~GeV.

Table 4: Events rates at $\sqrt{s_{e^+e^-}}=3$~TeV and $m_H=\infty$.
The invariant mass $M_{34}$ of central pair is required to be
500~GeV$<M_{34}<$1.5~TeV.

\newpage
\begin{table}
\caption{}
\begin{center}
\begin{tabular}{|c|c|c|c||c|c|c|} \hline
\multicolumn{1}{|c|}{$\sqrt{s_{e^+e^-}} =1.5$ TeV}
&\multicolumn{3}{|c||}{$\gamma\gamma\to WWWW$}
&\multicolumn{3}{|c|}{$\gamma\gamma\to WWZZ$} \\ \hline
 $m_H$, TeV & 0.1 & 0.5 & 0.7
& 0.1 & 0.5 & 0.7 \\ \hline
$\sigma_{tot}$, fb &
 30.6     & 40.4    & 34.7  &
 6.85    & 11.6    & 8.92
\\ \hline \hline
\multicolumn{1}{|c|}{$\sqrt{s_{e^+e^-}} =2$ TeV}
&\multicolumn{3}{|c||}{$\gamma\gamma\to WWWW$}
&\multicolumn{3}{|c|}{$\gamma\gamma\to WWZZ$} \\ \hline
 $m_H$, TeV & 0.1 & 1  & $\infty$
& 0.1 & 1 & $\infty$ \\ \hline
$\sigma_{tot}$, fb &
 61.1     & 68.1    & 65.6  &
 14.3     & 17.9    & 16.1
\\ \hline \hline
\multicolumn{1}{|c|}{$\sqrt{s_{e^+e^-}} =3$ TeV}
&\multicolumn{3}{|c||}{$\gamma\gamma\to WWWW$}
&\multicolumn{3}{|c|}{$\gamma\gamma\to WWZZ$} \\ \hline
 $m_H$, TeV & 0.1 & $\infty$ &
& 0.1 & $\infty$ &  \\ \hline
$\sigma_{tot}$, fb &
 133     &  147    &  &
 32.2    & 38.3    &
\\ \hline
\end{tabular}
\end{center}
\end{table}

\begin{table}
\caption{}
\begin{center}
\begin{tabular}{|c|c|c|c|c|c|c|c|c|} \hline
\multicolumn{9}{|c|}{$\sqrt{s_{e^+e^-}} = 2$ TeV;\quad $\int{\cal
L}dt=300$~fb$^{-1}$;\quad $m_H=1$~TeV;\quad $\Gamma_H=0.52$~TeV} \\ \hline
$WWWW$ &\multicolumn{4}{|c|}{$|\eta_{1,2}|>1.5$, $|\eta_{3,4}|<1$}
&\multicolumn{4}{|c|}{$|\eta_{3,4}|<1$}\\ \cline{2-9}
& $S$ & $B$ & $S/B$ & $S/\sqrt{B}$
& $S$ & $B$ & $S/B$ & $S/\sqrt{B}$ \\ \hline
---
&     65 &     78 & 0.83     &  7.3
&    149 &    453 & 0.33     &  7.0    \\
$|\eta_{1\div 4}|<3$
&     42 &     66 & 0.63     &  5.2
&    117 &    429 & 0.27     &  5.6    \\
$|\eta_{1\div 4}|<2.5$
&     22 &     50 & 0.44     &  3.1
&     86 &    389 & 0.22     &  4.3    \\ \hline \hline
$WWZZ$ &\multicolumn{4}{|c|}{$|\eta_{1,2}|>1.5$, $|\eta_{3,4}|<1$}
&\multicolumn{4}{|c|}{$|\eta_{3,4}|<1$}\\ \cline{2-9}
& $S$ & $B$ & $S/B$ & $S/\sqrt{B}$
& $S$ & $B$ & $S/B$ & $S/\sqrt{B}$ \\ \hline
---
   &     41 &     24 &  1.7     &  8.3
   &     78 &    165 & 0.47     &  6.1    \\
$|\eta_{1\div 4}|<3$
   &     25 &     20 &  1.3     &  5.7
   &     57 &    157 & 0.36     &  4.6    \\
$|\eta_{1\div 4}|<2.5$
   &     15 &     15 &  1.0     &  3.9
   &     40 &    146 & 0.28     &  3.3    \\ \hline \hline
{}~~~~$WWWW$ &\multicolumn{4}{|c|}{$|\eta_{1,2}|>1.5$, $|\eta_{3,4}|<1$}
&\multicolumn{4}{|c|}{$|\eta_{3,4}|<1$}\\ \cline{2-9}
$+\ WWZZ$ & $S$ & $B$ & $S/B$ & $S/\sqrt{B}$
& $S$ & $B$ & $S/B$ & $S/\sqrt{B}$ \\ \hline
---
   &    105 &    101 &  1.0     &  10
   &    227 &    617 & 0.37     &  9.1    \\
$|\eta_{1\div 4}|<3$
   &     67 &     86 & 0.78     &  7.3
   &    174 &    586 & 0.30     &  7.2    \\
$|\eta_{1\div 4}|<2.5$
   &     37 &     65 & 0.57     &  4.6
   &    126 &    535 & 0.24     &  5.4    \\ \hline \hline
\multicolumn{9}{|c|}{$\sqrt{s_{e^+e^-}} = 2$ TeV;\quad $\int{\cal
L}dt=300$~fb$^{-1}$;\quad $m_H=\infty$} \\ \hline
{}~~~~$WWWW$ &\multicolumn{4}{|c|}{$|\eta_{1,2}|>1.5$, $|\eta_{3,4}|<1$}
&\multicolumn{4}{|c|}{$|\eta_{3,4}|<1$}\\ \cline{2-9}
$+\ WWZZ$ & $S$ & $B$ & $S/B$ & $S/\sqrt{B}$
& $S$ & $B$ & $S/B$ & $S/\sqrt{B}$ \\ \hline
---
   &     33  &    101  & 0.33     &  3.3
   &     88  &    617  & 0.14     &  3.5    \\
$|\eta_{1\div 4}|<3$
   &     21  &     86  & 0.24     &  2.3
   &     67  &    586  & 0.11     &  2.8    \\
$|\eta_{1\div 4}|<2.5$
   &      8  &     65  & 0.13     &  1.0
   &     46  &    535  & 0.09 &  2.0    \\ \hline
\end{tabular}
\end{center}
\end{table}

\begin{table}
\caption{}
\begin{center}
\begin{tabular}{|c|c|c|c|c|c|c|c|c|} \hline
\multicolumn{9}{|c|}{$\sqrt{s_{e^+e^-}} = 1.5$ TeV;\quad $\int{\cal
L}dt=200$~fb$^{-1}$;\quad $m_H=500$~GeV;\quad $\Gamma_H=64$~GeV} \\ \hline
{}~~~~$WWWW$ &\multicolumn{4}{|c|}{$|\eta_{1,2}|>1.5$, $|\eta_{3,4}|<1$}
&\multicolumn{4}{|c|}{$|\eta_{3,4}|<1$}\\ \cline{2-9}
$+\ WWZZ$ & $S$ & $B$ & $S/B$ & $S/\sqrt{B}$
& $S$ & $B$ & $S/B$ & $S/\sqrt{B}$ \\ \hline
---
   &    122  &     29  &  4.2     &  23
   &    236  &    197  &  1.2     &  17     \\
$|\eta_{1\div 4}|<3$
   &     85  &     26  &  3.3     &  17
   &    186  &    189  & 0.98     &  14     \\
$|\eta_{1\div 4}|<2.5$
   &     50  &     20  &  2.5     &  11
   &    134  &    174  & 0.77     &  10     \\ \hline \hline
\multicolumn{9}{|c|}{$\sqrt{s_{e^+e^-}} = 1.5$ TeV;\quad $\int{\cal
L}dt=200$~fb$^{-1}$;\quad $m_H=700$~GeV;\quad $\Gamma_H=180$~GeV} \\ \hline
{}~~~~$WWWW$ &\multicolumn{4}{|c|}{$|\eta_{1,2}|>1.5$, $|\eta_{3,4}|<1$}
&\multicolumn{4}{|c|}{$|\eta_{3,4}|<1$}\\ \cline{2-9}
$+\ WWZZ$
& $S$ & $B$ & $S/B$ & $S/\sqrt{B}$
& $S$ & $B$ & $S/B$ & $S/\sqrt{B}$ \\ \hline
---
&   27  &     20  &  1.4     &  6.0 &     67  &    180  & 0.37  &  5.0\\
$|\eta_{1\div 4}|<3$
&     20  &     18  &  1.2     &  4.9 &     56  & 174  & 0.32   &  4.3\\
$|\eta_{1\div 4}|<2.5$ &     14  &     14  &
1.0     &  3.7 &     45  &    163  & 0.28     &  3.5    \\ \hline
\end{tabular}
\end{center}
\end{table}

\begin{table}
\caption{}
\begin{center}
\begin{tabular}{|c|c|c|c|c|c|c|c|c|} \hline
\multicolumn{9}{|c|}{$\sqrt{s_{e^+e^-}} = 3$ TeV;\quad $\int{\cal
L}dt=300$~fb$^{-1}$;\quad $m_H=\infty$} \\ \hline
{}~~~~$WWWW$ &\multicolumn{4}{|c|}{$|\eta_{1,2}|>1.5$, $|\eta_{3,4}|<1$}
&\multicolumn{4}{|c|}{$|\eta_{3,4}|<1$}\\  \cline{2-9}
$+\ WWZZ$ & $S$ & $B$ & $S/B$ & $S/\sqrt{B}$
& $S$ & $B$ & $S/B$ & $S/\sqrt{B}$ \\ \hline
---
   &    206  &    296  & 0.70     &  12
   &    335  &   1116  & 0.30     &  10     \\
$|\eta_{1\div 4}|<3$
   &     94  &    230  & 0.41     &  6.2
   &    194  &   1002  & 0.19     &  6.1    \\
$|\eta_{1\div 4}|<2.5$
   &     48  &    151  & 0.32     &  3.9
   &    120  &    855  & 0.14     &  4.1    \\ \hline
\end{tabular}
\end{center}
\end{table}

\end{document}